\documentclass{article} 
\usepackage{iclr2026_conference,times}

\usepackage{amsmath,amsfonts,bm}

\def\eqref#1{equation~\ref{#1}}

\def\1{\bm{1}}

\DeclareMathAlphabet{\mathsfit}{\encodingdefault}{\sfdefault}{m}{sl}
\SetMathAlphabet{\mathsfit}{bold}{\encodingdefault}{\sfdefault}{bx}{n}

\usepackage{hyperref}
\hypersetup{hidelinks}
\usepackage{url}
\usepackage{amsmath,amssymb,amsthm}
\usepackage{booktabs}
\usepackage{caption}
\usepackage{multirow}
\usepackage{graphicx}
\usepackage{bm}
\usepackage{tikz}
\usetikzlibrary{arrows.meta,positioning,fit,backgrounds,calc}
\usepackage[ruled,vlined]{algorithm2e}
\newtheorem{proposition}{Proposition}
\newtheorem{assumption}{Assumption}
\newtheorem{remark}{Remark}

\newcommand{\ket}[1]{\lvert #1 \rangle}
\newcommand{\bra}[1]{\langle #1 \rvert}

\usepackage{xcolor}

\title{Fixed-Protocol Amortized MPS Tomography\\ with Conformalized Predictive Uncertainty}

\author{Jian Xu$^{1,2}$, \; Delu Zeng$^{3}$, \; John Paisley$^{4}$, \; Qibin Zhao$^{2}$ \\[2pt]
$^{1}$RIKEN iTHEMS \quad $^{2}$RIKEN AIP \quad $^{3}$South China University of Technology \quad
$^{4}$Columbia University \\[2pt]
\texttt{jian.xu@riken.jp}}

\iclrfinalcopy 
\begin{document}

\maketitle
\lhead{}\chead{}\rhead{}\renewcommand{\headrulewidth}{0pt} 
\thispagestyle{fancy}

\begin{abstract}
Quantum state tomography is sample-starved, and the states one prepares live on a narrow, learnable
manifold. A $k{=}0$ prior-only control shows that on concentrated families a prior estimate is already
near-optimal, so ``high fidelity at few measurements'' can be family memorization rather than tomography;
genuine measurement-efficiency needs a model that conditions on the measurements and demonstrably uses them.
On a shared matrix-product-state (MPS) core parameterization we study two routes. \emph{Approach~A} learns a
generative prior over MPS cores with measurement-guided posterior inference (gold-standard-validated, but
whose few-measurement accuracy the control shows is largely the prior). \emph{Approach~B}, our main
proposal, is a \emph{fixed-protocol amortized} MPS estimator trained once with a gauge-invariant fidelity
loss; we deliberately do not rest it on a permutation-invariant set encoder (a plain MLP matches it). The
decisive lever is the \emph{measurement design}: motivated by the fact that local reduced density matrices
determine a $\chi$-MPS, conditioning on an \emph{informative local} Pauli set rather than random strings
turns a modest, memorization-prone estimator into a high-fidelity one ($\approx\!0.95$, up to $+0.59$ over
prior-only, decisively passing a shuffled-measurement control). A dropout ensemble, conformally
recalibrated, gives $\approx\!90\%$-coverage intervals---including for observables never measured, where a
shot-based interval does not exist. Quality holds as the system grows (fidelity $0.90$ at $n{=}10$, gain
\emph{growing} in $n$; $0.88$ at bond dimension $\chi{=}4$), the parameterization is polynomial (native
contraction to $20$ qubits), and we close the loop on IBM hardware ($5$ states at $0.97$ from
hardware-measured Paulis). Our prior-only and shuffled controls are a necessary integrity check we argue
learned-QST work should adopt.
\end{abstract}

\section{Introduction}
Quantum state tomography (QST)---estimating an unknown quantum state $\rho$ from measurement
data---is a cornerstone primitive for benchmarking and debugging quantum hardware. Its central
difficulty is \emph{sample complexity}: full tomography of an $n$-qubit state requires a number
of measurements that scales exponentially in $n$, and each measurement (state preparation plus
readout) is expensive. Reducing the measurement budget needed to reach a target fidelity is
therefore the practically important axis.

Two broad families reduce this cost. \emph{Structural} methods restrict the model class:
compressed-sensing QST exploits low rank \citep{gross2010quantum}, and matrix-product-state (MPS)
tomography exploits low entanglement \citep{cramer2010efficient, lanyon2017efficient}, both recovering
poly-parameter states from poly-many measurements. \emph{Bayesian} methods
\citep{blume2010optimal, granade2016practical, lukens2020practical} return a posterior---and hence uncertainty---but
rely on hand-chosen priors on the dense state and do not learn from the distribution of states a
device actually produces. Neither family uses the key empirical fact that the states prepared in
a given experiment are not arbitrary: they lie on a \emph{narrow, structured manifold} (e.g.
ground states of a Hamiltonian family, or the output states of a parameterized circuit).

\paragraph{A learned prior alone is not enough---and a control that shows it.}
The tempting recipe is to learn a generative \emph{prior} over the family and invert by posterior
sampling \citep{chung2022diffusion}. We build this (Approach~A: a flow-matching prior over MPS cores with a
measurement-guided proximal/Doob-$h$-transform sampler)---a genuine, gold-standard-validated Bayesian
posterior---and it produces high fidelity at few measurements. But a simple control deflates the
interpretation. On a concentrated family---e.g.\ disordered paramagnetic ground states, which are all
close to $\ket{+}^{\otimes n}$---a \emph{prior-only} estimate that uses \emph{zero} measurements (the
family medoid) already reaches fidelity $0.997$, and \emph{shuffling} the measurements (reconstructing
from another state's data) changes nothing. There, ``measurement-efficient tomography'' is really
nearest-neighbor prediction from a narrow distribution. The lesson is methodological: a learned-QST method
must be judged against a $k{=}0$ prior-only baseline and must \emph{use} the measurements, verifiable by a
shuffled-measurement test. We adopt both as integrity controls throughout.

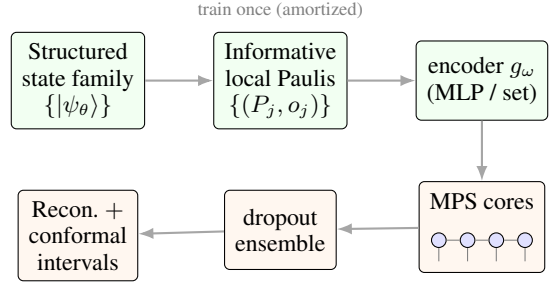
\begin{figure}[t]
\centering
\begin{tikzpicture}[
  font=\small,
  box/.style={draw,rounded corners=2pt,align=center,inner sep=4pt,minimum height=1.05cm,fill=blue!4},
  train/.style={box,fill=green!5},
  infer/.style={box,fill=orange!6},
  ar/.style={-{Latex[length=2mm]},thick,gray!70},
  site/.style={circle,draw,fill=blue!12,minimum size=5.5pt,inner sep=0pt}]
\newcommand{\mpsdiag}[1]{\begin{tikzpicture}[baseline=-2pt]
  \foreach \x in {0,1,2,3}{\node[site] (a\x) at (0.38*\x,0){}; \draw[gray] (a\x)--++(0,-0.28);}
  \draw[gray] (a0)--(a1)--(a2)--(a3);\end{tikzpicture}}
\node[train] (fam) {Structured\\state family\\ $\{\ket{\psi_\theta}\}$};
\node[train,right=0.9cm of fam] (meas) {Informative\\local Paulis\\ $\{(P_j,o_j)\}$};
\node[train,right=0.9cm of meas] (enc) {encoder $g_\omega$\\ (MLP / set)};
\draw[ar] (fam)--(meas); \draw[ar] (meas)--(enc);
\node[above=0.05cm of meas,gray,font=\scriptsize] {train once (amortized)};
\node[infer,below=0.8cm of enc] (cores) {MPS cores\\[1pt] \mpsdiag{}};
\node[infer,below=0.8cm of meas] (mc) {dropout\\ ensemble};
\node[infer,below=0.8cm of fam] (out) {Recon.\ $+$\\ conformal\\ intervals};
\draw[ar] (enc)--(cores); \draw[ar] (cores)--(mc); \draw[ar] (mc)--(out);
\end{tikzpicture}
\caption{Fixed-protocol amortized tomography. A fixed-protocol encoder $g_\omega$ (a plain MLP or a
permutation-invariant set encoder---the architecture is not the source of performance) maps the measured
values $\{(P_j,o_j)\}$---at a fixed informative local Pauli design---directly to matrix-product-state cores,
trained once on a state family with a gauge-invariant fidelity loss. A dropout predictive ensemble,
recalibrated conformally, yields a reconstruction and calibrated observable intervals.}
\label{fig:overview}
\end{figure}

\paragraph{Conditioning on measurements, with the right measurement design.}
To genuinely use the data, we replace the unconditional prior$+$inference pipeline with an
\emph{amortized, measurement-conditioned estimator}: a fixed-protocol encoder (a plain MLP or a permutation-invariant set-encoder) reads the measured pairs
$\{(P_j,o_j)\}$ and outputs the cores of an MPS, trained once on the family with a gauge-invariant
state-fidelity loss (Fig.~\ref{fig:overview}). Conditioning is necessary but not sufficient: with
\emph{random} Pauli strings the conditional model still only modestly beats prior-only, because random
high-weight Paulis are uninformative about low-entanglement states. The decisive lever is the
\emph{measurement design}. A classical result---local reduced density matrices determine a $\chi$-MPS
\citep{cramer2010efficient}---says the informative observables are \emph{local} (Prop.~\ref{prop:localdesign}). We condition on a
cheap local set (weight-$1$ and nearest-neighbor weight-$2$ Paulis), and also test the theory-complete
$3$-site-window design; both work, whereas random Paulis do not.
Switching from random to informative-local measurements turns a modest, memorization-prone estimator into
a high-fidelity one that clearly beats prior-only and passes the shuffled control.

\paragraph{Why cores.} The MPS-core parameterization is load-bearing: a generative model over the dense
$2^n$ state vector underfits already at $n=6$, whereas the conditional model over $O(n\chi^2)$ cores
learns the family and reconstructs it; and the transfer-matrix contraction is polynomial, so the pipeline
runs to $20$ qubits without ever forming the state vector. MPS cores carry a gauge (a bond-wise change of
basis); we fix a deterministic coordinatization (left-canonical QR) but do not oversell it---it is a
coordinate choice, not a state-level invariant (Prop.~\ref{prop:coorddep}), and we train the model with a
gauge-invariant target so the choice does not bias reconstruction.

\paragraph{Contributions.}
\begin{itemize}
    \item \textbf{Integrity controls for learned QST.} A $k{=}0$ prior-only baseline and a
    shuffled-measurement test, showing a learned prior can look good without doing tomography, and that
    these controls are necessary. Approach~B passes both; the prior-only/unconditional route does not.
    \item \textbf{A generative prior over tensor-network cores (Approach~A).} A flow-matching prior over
    canonicalized MPS cores with measurement-guided (Doob $h$-transform / proximal) Bayesian posterior
    inference---to our knowledge the first learned generative distribution over tensor-network cores---%
    validated against a gold-standard importance sampler. The integrity controls show its few-measurement
    accuracy leans on the prior, which motivates Approach~B.
    \item \textbf{A fixed-protocol amortized MPS estimator}: a network mapping one fixed informative
    measurement protocol $\to$ tensor-network cores, trained once with a gauge-invariant loss, with a
    dropout-ensemble predictive distribution recalibrated to distribution-free marginal coverage for the
    measured observables. To our knowledge this is the first amortized measurement-conditioned
    \emph{full-state} MPS estimator for QST (vs.\ per-state fits \citep{torlai2023quantum}, property
    point-estimators \citep{cha2025scalable}, or uncalibrated generative reconstructions \citep{ahmed2021quantum}).
    We are deliberate about what does the work: a controlled ablation shows a \emph{plain MLP} on the
    fixed-order values matches a permutation-invariant set-encoder, so permutation-invariance is \emph{not}
    the source of performance---the contribution is the fixed informative protocol, the MPS-core target, and
    the gauge-invariant loss, not the encoder architecture. We also do \emph{not} call the dropout ensemble a
    Bayesian posterior.
    \item \textbf{An informative local measurement design}, grounded in a constructive spanning-design
    proposition, identified as the lever that makes conditioning high-fidelity (a $0.79\!\to\!0.95$
    fidelity, $+0.12\!\to\!+0.59$ prior-only-gain swing from random to local Paulis).
    \item \textbf{Theory} (a formalization, not the main contribution): coordinate-dependence of core
    priors (Prop.~\ref{prop:coorddep}); a constructive $O(n)$ local spanning design
    (Prop.~\ref{prop:localdesign}); the structured-vs-unstructured degrees-of-freedom separation; and
    distribution-free conformal coverage.
    \item \textbf{Scalability and hardware}: reconstruction quality that holds as $n$ grows (fidelity $0.90$
    at $n{=}10$ with the measurement gain growing in $n$), native $O(n\chi^3)$ transfer-matrix contraction to
    $20$ qubits, and a \emph{closed-loop} reconstruction on IBM hardware ($5$ states at fidelity $0.97$ from
    hardware-measured local Paulis).
\end{itemize}

\section{Related Work}
\label{sec:related}
We position against the nearest neighbors along each axis; the point is that each pairwise
combination is occupied but the intersection is not.

\paragraph{Structural / MPS tomography.} MPS-QST \citep{cramer2010efficient, lanyon2017efficient} and
tensor-train compressed-sensing QST \citep{sofi2025tensor} recover a state by \emph{fitting} cores to
data under a hand-designed low-rank/sparsity prior. \citet{torlai2023quantum} learn a tensor-network
channel by gradient descent to fit measurement statistics. In all of these the cores are the
optimization variables for a \emph{single} unknown state; none learns a transferable generative
distribution over cores. Our prior is exactly such a distribution, and it is what makes the
few-measurement regime tractable.

\paragraph{Generative / neural QST.} Neural QST fits a generative model (RBM, autoregressive,
transformer) to the measurements of \emph{one} target state
\citep{torlai2018neural, carrasquilla2019reconstructing, cha2022attention}; QST-CGAN \citep{ahmed2021quantum} amortizes a
GAN across similar states but returns a point estimate. Diffusion has been used to \emph{generate}
valid density matrices \citep{zhu2024quantum}, not to reconstruct from measurements. Closest
in spirit, QuaDiM \citep{tang2025quadim} is a conditional diffusion model in the quantum-measurement
space, but it conditions on \emph{Hamiltonian parameters} (not on measurements of an unknown state) and
\emph{generates} measurement snapshots for property estimation rather than inverting given measurements
to reconstruct a state, and it uses no tensor-network representation. We differ by learning an amortized
conditional \emph{estimator} of MPS cores directly from a fixed informative Pauli protocol on an unknown
state, with distribution-free calibrated observable intervals.

\paragraph{Amortized / learned-prior Bayesian QST.} Closest to our motivation,
\citet{lohani2023demonstration} learn (with a shallow CNN) the hyperparameters of a Bayesian prior; and
\citet{cha2025scalable} amortize a set-transformer estimate of the Bayesian posterior \emph{mean} of
state \emph{properties} from classical shadows. Both are point estimators (of properties or of
prior hyperparameters); neither learns an amortized conditional full-state \emph{estimator} over
tensor-network cores nor reports calibrated, distribution-free observable-level intervals for the
reconstructed state. Simulation-based inference with normalizing-flow
posteriors \citep{papamakarios2016fast} and coverage calibration \citep{falkiewicz2023calibrating}
are mature outside physics but, to our knowledge, unapplied to QST.

\paragraph{Generative priors for inverse problems (Approach~A).} A large body of work solves inverse
problems by sampling a data-consistent posterior under a learned generative prior---diffusion/flow
posterior sampling and its guidance schemes \citep{chung2022diffusion, song2023pseudoinverse, wu2023practical},
and their formalization through Doob $h$-transforms and twisted/guided diffusion
\citep{denker2024deft, wu2023practical}. These are developed for images and general Bayesian inverse
problems; our Approach~A instantiates this program in QST by placing a rectified-flow prior on
\emph{canonicalized MPS cores} and guiding it with the Pauli shot-likelihood, then validating against a
gold-standard SIR posterior. The novel ingredients are the tensor-network core parameterization (which
makes the prior learnable) and the integrity controls that reveal when the prior, rather than the
measurements, is doing the work.

\paragraph{Uncertainty in QST.} Rigorous frequentist confidence regions
\citep{christandl2012reliable, faist2016practical} and, recently, anytime-valid confidence sequences
\citep{cumitini2025anytime} set the calibration bar but use no learned prior. Bayesian credible-region
methods \citep{blume2010optimal, granade2016practical, oh2019efficient} use generic priors. The
size-growing advantage of \emph{structured} priors is known for point estimation
\citep{cramer2010efficient}; our contribution is to realize and quantify it in a \emph{learned, calibrated,
core-parameterized} setting.

\section{Background}
\paragraph{Setup.} Let $\ket{\psi}\in\mathbb{C}^{2^n}$ be an $n$-qubit pure state. We observe noisy
estimates of Pauli expectation values $o_j = \bra{\psi}P_j\ket{\psi} + \varepsilon_j$ for a set of
Pauli strings $\{P_j\}_{j=1}^k$, with shot noise $\varepsilon_j$ of variance
$(1-\langle P_j\rangle^2)/S$ for $S$ shots per setting. We keep $S{=}200$ fixed throughout, so the total
measurement cost is $kS$ state copies; the free axis we sweep is the number of distinct Pauli settings
$k$ (each an independent measurement configuration), and we report budgets as $k$ at fixed $S$ to avoid
conflating settings with copies. The goal is to estimate $\ket{\psi}$ and conformalized observable
intervals, ideally with $k \ll 4^n$.

\paragraph{MPS and gauge.} An MPS with bond dimension $\chi$ represents amplitudes as
$\psi(s_1\dots s_n) = A^{(1)}_{s_1}A^{(2)}_{s_2}\cdots A^{(n)}_{s_n}$, with cores
$A^{(\ell)}\in\mathbb{C}^{\chi_{\ell-1}\times 2\times \chi_\ell}$ ($\chi_0=\chi_n=1$). The
representation is invariant under bond-wise gauge transformations
$A^{(\ell)}\!\to\!A^{(\ell)}G_\ell,\; A^{(\ell+1)}\!\to\!G_\ell^{-1}A^{(\ell+1)}$. A generative model
over cores must therefore act on a \emph{canonical form}; we use the left-canonical gauge (successive
QR with a sign convention on the triangular factors), a \emph{norm normalization} of the boundary
core, and a global-phase fix. This is the tensor-network analogue of removing permutation/sign/scaling
ambiguity in matrix/CP factorizations. The normalization is not cosmetic: without it the state norm
concentrates into a single core entry, making the standardized core coordinates span orders of
magnitude ($\sim\!10^4$ at $n{=}10$) and defeating the flow; normalizing the boundary core (the
left-canonical cores are isometries, so the norm equals the boundary core's Frobenius norm) keeps all
coordinates $O(1)$ and is what makes the prior learnable at scale (Sec.~\ref{sec:exp}).

\begin{figure}[t]
\centering
\begin{tikzpicture}[font=\footnotesize,x=1.05cm,y=0.62cm,
  g/.style={draw,fill=blue!8,rounded corners=1.5pt,minimum width=0.62cm,minimum height=0.42cm,inner sep=1pt},
  m/.style={draw,fill=orange!12,rounded corners=1.5pt,minimum width=0.55cm,minimum height=0.42cm,inner sep=1pt}]
\foreach \q in {0,1,2,3}{\draw[gray] (0,-\q)--(6.2,-\q); \node[left] at (-0.05,-\q) {$q_{\q}$};}
\foreach \q in {0,1,2,3}{\node[g] at (0.7,-\q) {$R_y$};}
\foreach \q in {0,1,2}{\pgfmathtruncatemacro\t{\q+1}
  \filldraw (1.8,-\q) circle (1.6pt); \draw (1.8,-\q)--(1.8,-\t); \draw (1.8,-\t) circle (4pt);}
\foreach \q in {0,1,2,3}{\node[g] at (3.0,-\q) {$R_y$};}
\draw[dashed,gray!60] (3.9,0.5) -- (3.9,-3.5);
\node[gray] at (2.0,0.9) {structured state $\ket{\psi_\theta}$};
\foreach \q in {0,1,2,3}{\node[m] at (4.9,-\q) {$P_j$};}
\node[gray] at (5.6,0.9) {local Pauli};
\foreach \q in {0,1,2,3}{\draw[-{Latex[length=1.4mm]}] (5.25,-\q)--(6.1,-\q);}
\end{tikzpicture}
\caption{The structured state family and measurement model. A shallow $R_y$--CNOT--$R_y$ circuit with
random angles prepares a low-entanglement state $\ket{\psi_\theta}$; noisy expectation values of
the informative-local Pauli observables $P_j$ (single-qubit and nearest-neighbor) constitute the few-measurement tomographic data.}
\label{fig:circuit}
\end{figure}
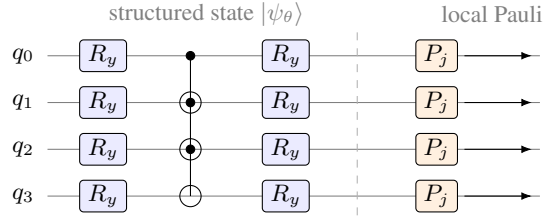

\paragraph{Structured state family.} A running physical example is generated by a shallow
parameterized circuit (Fig.~\ref{fig:circuit}): a layer of single-qubit rotations $R_y(\theta)$, a
nearest-neighbour entangling ladder, and a second rotation layer, with random angles---yielding a
low-dimensional, low-entanglement manifold of states represented exactly at bond dimension
$\chi{=}2$. This stands in for physically-relevant families (e.g.\ Hamiltonian ground states), and its
angles set the family's diversity.

\section{Method}
\label{sec:method}

\paragraph{Canonicalized core coordinates.}
We map each state $\ket{\psi}$ of the family to a canonical core vector as follows. Sweeping
$\ell=1,\dots,n{-}1$, reshape the $\ell$-th core to
$A^{(\ell)}\!\in\!\mathbb C^{(\chi_{\ell-1}2)\times\chi_\ell}$ and take a (thin) QR factorization
\begin{equation}
A^{(\ell)}=Q^{(\ell)}R^{(\ell)},\quad
S^{(\ell)}=\mathrm{diag}\!\big(\mathrm{sgn}\,R^{(\ell)}_{ii}\big),\qquad
A^{(\ell)}\!\leftarrow\! Q^{(\ell)}S^{(\ell)},\;\;
A^{(\ell+1)}\!\leftarrow\! S^{(\ell)}R^{(\ell)}A^{(\ell+1)},
\label{eq:leftcanon}
\end{equation}
so every $A^{(\ell)}$ ($\ell<n$) is a left isometry, $A^{(\ell)\dagger}A^{(\ell)}=I$, and the diagonal
sign fix removes the residual bond phase. We then normalize the boundary core and fix the global phase,
\begin{equation}
A^{(n)}\leftarrow A^{(n)}/\lVert A^{(n)}\rVert_F,\qquad
A^{(n)}\leftarrow e^{-i\arg\psi_{0}(\bm\theta)}A^{(n)},
\label{eq:normphase}
\end{equation}
where $\psi_0$ is the first amplitude; because the left cores are isometries,
$\lVert\psi(\bm\theta)\rVert=\lVert A^{(n)}\rVert_F$, so \eqref{eq:normphase} yields a unit-norm state
and keeps all coordinates $O(1)$ (Sec.~\ref{sec:exp}). Stacking the real and imaginary parts of the
canonical cores and standardizing gives $\bm\theta\in\mathbb R^{d_\theta}$, $d_\theta=O(n\chi^2)$.

\paragraph{Two approaches, one parameterization.} On this shared core parameterization we study two ways to
turn measurements into a reconstruction, unified by the integrity controls of Sec.~\ref{sec:exp}.
\emph{Approach~A} learns an unconditional generative \emph{prior} over cores and inverts it by
measurement-guided posterior sampling: it yields a genuine, gold-standard-validated Bayesian posterior, but
the controls reveal it leans heavily on the prior. \emph{Approach~B} conditions on the measurements
directly, extracting more measurement information and---with the informative-local design---reconstructing
at high fidelity. A and B share the MPS-core parameterization, the gauge treatment, the native contraction,
and the conformal calibration; they differ in whether the measurements enter through a likelihood at
inference (A) or through the network's input (B).

\paragraph{Approach~A: generative prior and Bayesian posterior inference.} We train a rectified-flow prior
$v_\phi$ over standardized canonical cores $\bm\theta$ by flow matching,
$\mathcal L(\phi)=\mathbb E_{\bm\theta,\bm\epsilon,t}\lVert v_\phi(\bm\theta_t,t)-(\bm\theta-\bm\epsilon)\rVert^2$
with $\bm\theta_t=(1{-}t)\bm\epsilon+t\bm\theta$. This is, to our knowledge, the first learned
\emph{generative} distribution over tensor-network cores (as opposed to a tensor network used \emph{as} a
generative model \citep{han2018unsupervised} or fit to a single state \citep{torlai2023quantum}). Given data $\bm o$ with
the heteroscedastic shot-noise likelihood
$\ell(\bm o\mid\bm\theta)\propto\exp(-\tfrac12\sum_j(\langle P_j\rangle_{\bm\theta}-o_j)^2/\varsigma_j^2)$,
$\varsigma_j^2=(1{-}\langle P_j\rangle^2)/S$, the posterior-conditioned flow is the Doob $h$-transform of the
prior; we approximate its intractable guidance by a proximal data-consistency step on the flow's clean-core
estimate at each integration step (a DiffPIR/DPS-style guided sampler \citep{chung2022diffusion}), producing
posterior samples and a reconstruction on the low-dimensional core manifold. This is a principled,
measurement-guided posterior sampler---we validate it against a gold-standard importance sampler in
Sec.~\ref{sec:exp}---and it is the ``unconditional-prior'' route the integrity controls scrutinize.

\paragraph{Approach~B: amortized, measurement-conditioned estimator.}
Rather than learn an unconditional prior and invert it, we learn the map from measurements to state
directly. Fix a measurement design---an ordered set of $k$ Pauli observables $\mathcal D=\{P_j\}$ (we argue
below this should be an \emph{informative local} set). Given noisy expectations $\bm o=(o_j)$ with
$o_j=\langle P_j\rangle_{\psi_\star}+\varepsilon_j$, we learn a direct amortized \emph{estimator}
$g_\omega:\{(P_j,o_j)\}_{j=1}^k\mapsto\bm\theta$ of the state's canonical cores (a point reconstruction; a
calibrated predictive spread is added post hoc in \S\ref{sec:exp}), rather than sampling a posterior. For a
\emph{fixed} protocol the encoder architecture is not what matters (a plain MLP on the fixed-order value
vector matches a set-encoder; \S\ref{sec:exp}); by default we use a permutation-invariant set-encoder so
the same code also accepts variable designs: each pair is embedded as
$h_j=\mathrm{MLP}\big([\mathrm{onehot}(P_j);\,o_j]\big)$, pooled permutation-invariantly
$z=\tfrac1k\sum_j h_j$, and decoded to cores $\bm\theta=\mathrm{MLP}(z)$ (App.~\ref{app:details}).

\paragraph{Gauge-invariant training.}
We train $g_\omega$ on (state, measurement) pairs from the family, minimizing a \emph{state-space} loss
\begin{equation}
\mathcal L(\omega)=\mathbb E_{\ket{\psi}\sim\text{family},\,\bm o\mid\ket\psi}\Big[\,1-\big|\langle\psi\,|\,\psi(g_\omega(\bm o))\rangle\big|^2\,\Big],
\label{eq:condloss}
\end{equation}
one minus reconstruction fidelity. This is deliberately \emph{gauge-invariant}: since the MPS gauge and
global phase leave $\ket{\psi(\bm\theta)}$ unchanged, \eqref{eq:condloss} never penalizes the network for
choosing a different core representative of the correct state. A core-space regression loss
$\lVert g_\omega(\bm o)-\bm\theta_\star\rVert^2$ is \emph{not} gauge-invariant: while each state has a
single deterministic canonical target (so this is not a literal multi-label conflict), the canonicalization
is a \emph{discontinuous} coordinatization in which Euclidean core distance does not track state fidelity
(Prop.~\ref{prop:coorddep})---nearby states can have far-apart cores near spectral degeneracies. A core-MSE
target is therefore geometrically mismatched to the quantity we care about, and we find empirically that it
collapses whereas the fidelity loss does not (Sec.~\ref{sec:exp}). The
canonicalization \eqref{eq:leftcanon}--\eqref{eq:normphase} only defines the coordinate the network
predicts; the loss and the physical readout are gauge-invariant.

\paragraph{The measurement design is the lever: informative local Paulis.}
Conditioning is necessary, but the design $\mathcal D$ decides whether it helps. Random Pauli strings are
a poor design for low-entanglement states: a random weight-$w$ Pauli has expectation exponentially small
in $w$, so it carries little information and the conditional model barely beats prior-only. The
informative observables are \emph{local}. The precise statement is that local reduced density matrices on
windows of $\ell=2\lceil\log_2\chi\rceil{+}1$ sites determine a normal $\chi$-MPS \citep{cramer2010efficient}
(Prop.~\ref{prop:localdesign})---for $\chi{=}2$, tomographically complete Paulis on each contiguous
$3$-site window. This theory-complete design (\texttt{local3}) is one option we test. Our default
$\mathcal D$ is a cheaper subset---all weight-$1$ Paulis $\{X_i,Y_i,Z_i\}$ and weight-$2$ nearest-neighbor
Paulis $\{P_iP_{i+1}\}$ ($O(n)$)---which is \emph{not} proven complete but is empirically sufficient on our
families (Sec.~\ref{sec:exp}); the theory-complete window design does slightly better. Either way, the
point is that \emph{locality} is the lever: switching from random to local Paulis is what turns a modest
estimator into a high-fidelity one.

\paragraph{Predictive uncertainty and conformal calibration.}
The estimator $g_\omega$ is a point map; to attach uncertainty we use it as a \emph{dropout ensemble}:
with dropout active at test time, $E$ stochastic passes give predictions $\{\ket{\psi^{(e)}}\}$ and, for
each observable $O$, an ensemble mean $\mu_O$ and scale $\sigma_O$. We stress this is a heuristic
predictive spread, not a validated Bayesian posterior. The intervals $[\mu_O\pm\sigma_O]$ are
over-confident, so we \emph{recalibrate conformally}: on a held-out split we form scores
$r=|O-\mu_O|/\sigma_O$ and take the $\lceil(1-\alpha)(m{+}1)\rceil$-th smallest, $Q$, giving
$\mathcal I_O=[\mu_O\pm Q\sigma_O]$ with distribution-free \emph{marginal} coverage $\ge1-\alpha$ for that
fixed observable, under exchangeability of calibration and test states (Prop.~\ref{prop:cov}). This is
marginal per-observable coverage---not simultaneous coverage over observables, a full-state confidence
region, or coverage under distribution shift---and the same recalibration could be applied to any point
predictor, including the prior-only baseline; we report it for the conditional estimator because it is the
one with informative intervals.

\section{Theoretical analysis}
\label{sec:theory}
We give results that make the design choices precise: the physical readout and training objective are
gauge-invariant, but the core coordinatization---and hence the learned prior over cores---is not
(Prop.~\ref{prop:gauge}, Prop.~\ref{prop:coorddep}); local reduced density matrices give a
\emph{constructive} $O(n)$ informative measurement design (Prop.~\ref{prop:localdesign}), which is the
design our conditional model uses; the degrees of freedom of a manifold-restricted family separate
exponentially from unstructured tomography (Prop.~\ref{prop:adv}); and the recalibrated observable
intervals carry a distribution-free finite-sample coverage guarantee (Prop.~\ref{prop:cov}). Proofs are in
Appendix~\ref{app:proofs}.

Write $\Phi:\Theta\to\mathbb{C}^{2^n}$ for the MPS contraction $\bm\theta\mapsto\ket{\psi(\bm\theta)}$,
and let $\mathcal M=\{[\Phi(\bm\theta)]:\bm\theta\in\Theta\}$ be the induced family of physical states
(rays). The MPS gauge group acts on cores by
$A^{(\ell)}\!\mapsto\!A^{(\ell)}G_\ell,\,A^{(\ell+1)}\!\mapsto\!G_\ell^{-1}A^{(\ell+1)}$, leaving
$\Phi$ invariant. The Pauli-measurement map is
$m([\psi])=(\bra{\psi}P_j\ket{\psi})_{j=1}^k\in\mathbb R^k$.

\begin{proposition}[Gauge-invariant objective; well-definedness under a fixed coordinatization]
\label{prop:gauge}
(i) The training objective \eqref{eq:condloss} and the physical readout $\Phi$ depend on the cores only
through the physical state $[\Phi(\bm\theta)]$, and are therefore invariant to the MPS gauge and global
phase. (ii) For a fixed deterministic coordinatization $c$ (\eqref{eq:leftcanon}--\eqref{eq:normphase}),
applied consistently to define the network's target and to canonicalize its output, the reconstruction
map $\bm o\mapsto\Phi(c(g_\omega(\bm o)))$ is well-defined. We make no claim that this map is invariant
under a change of coordinatization $c\to c'$ (Prop.~\ref{prop:coorddep}); (i) only guarantees that the
objective and readout do not depend on \emph{which} gauge representative encodes a given state.
\end{proposition}

\begin{proposition}[Core coordinates are gauge-dependent; why we use a gauge-invariant loss]
\label{prop:coorddep}
Let the MPS gauge group $\mathcal G$ act on cores by state-preserving transformations
$\bm\theta\mapsto g\!\cdot\!\bm\theta$ (so $\Phi(g\!\cdot\!\bm\theta)=\Phi(\bm\theta)$). The
canonicalization $c$ is \emph{not} $\mathcal G$-equivariant: there is a positive-measure set of pairs
$\bm\theta,\bm\theta'=g\!\cdot\!\bm\theta$ with $\Phi(\bm\theta)=\Phi(\bm\theta')$ yet
$c(\bm\theta)\neq c(\bm\theta')$. Consequently any \emph{coordinate-space} objective---a density $q(\cdot)$
over canonical cores, or a core-regression loss $\lVert g_\omega(\bm o)-c(\bm\theta_\star)\rVert^2$---is not
a function of the physical state. We stress the \emph{practical} reading: in training each state yields a
\emph{single} deterministic canonical target, so this is not a literal multi-label conflict where one input
carries many gauge labels. Rather, because $c$ is discontinuous and Euclidean core distance does not track
state fidelity, a core-regression target is geometrically mismatched to the reconstruction objective; the
\emph{gauge-invariant} fidelity loss \eqref{eq:condloss} removes the mismatch, which is why we use it.
\end{proposition}
\begin{remark}
This is a statement about coordinates, and it motivates our gauge-invariant loss: empirically a
core-regression variant is markedly worse (\S\ref{sec:exp}). It is also why we say
``canonicalized'' rather than ``gauge-fixed''; a quotient-aware (e.g.\ Schmidt-coordinate) parameterization
is the principled alternative we leave open.
\end{remark}

\begin{proposition}[A constructive spanning design from local reduced density matrices]
\label{prop:localdesign}
Consider a site-dependent (not necessarily translation-invariant) \emph{normal} (injective) open-boundary
$\chi$-MPS with injectivity length $\ell_0$, and set $\ell=\max(2\ell_0{-}1,\,2\lceil\log_2\chi\rceil{+}1)$
(for the generic injective $\chi{=}2$ family here, $\ell_0{=}1$ and $\ell{=}3$). \citet[Thm.~2 and its
open-boundary reconstruction]{cramer2010efficient} show such a state is the \emph{unique} state---generically among
all states, not merely among MPS---consistent with its reduced density matrices on all contiguous
$\ell$-site windows. Each such reduced density matrix is a linear image of the expectations of the
weight-$\le\ell$ local Pauli strings supported on that window, giving an explicit, \emph{non-random} set of
$O(n\,4^{\ell})=O(n)$ Pauli observables. Wherever this local reconstruction is injective, the differentials
of this set span $T_{[\psi]}\mathcal M$, so it is a spanning design in the sense of
Assumption~\ref{ass:immersion}, giving identifiability from $k=O(n)$ measurement settings---versus
$2^{n+1}-2$ unstructured. We use this as \emph{motivation} for the design; it is a uniqueness/spanning
statement, not a noise-robust sample-complexity bound.
\end{proposition}

\begin{assumption}[Immersed family with a spanning Pauli design]
\label{ass:immersion}
$\mathcal M$ is a smooth embedded submanifold of the pure-state ray space of real dimension
$d=\dim\mathcal M\le 2\,c(n,\chi)$, where $c(n,\chi)=\sum_\ell \chi_{\ell-1}\cdot 2\cdot\chi_\ell = O(n\chi^2)$
is the number of complex core parameters; at the state of interest the measurement map $m|_{\mathcal M}$
is an immersion (its differential is injective on the tangent space); and the chosen set of $k$ Pauli
observables \emph{spans the cotangent image}, i.e.\ the differentials $\{\mathrm d\langle P_j\rangle\}_{j=1}^k$
have rank $d$ on $T_{[\psi]}\mathcal M$. (We assume such a spanning design \emph{exists} and is used; we do
\emph{not} claim that $d$ Pauli strings drawn uniformly at random attain rank $d$ with high probability.)
\end{assumption}

\begin{proposition}[Degrees-of-freedom separation, structured vs.\ unstructured]
\label{prop:adv}
Under Assumption~\ref{ass:immersion}, a spanning set of $k\ge d$ Pauli measurements makes $m|_{\mathcal M}$
a local diffeomorphism onto its image, so it locally identifies a state \emph{within} $\mathcal M$ from $d$
real numbers. In contrast, identifying an arbitrary pure state requires at least $2^{\,n+1}-2$ real
measurements (its intrinsic dimension). Hence the ratio of \emph{degrees of freedom} to be resolved is
$\;(2^{\,n+1}-2)/d=\Omega\!\left(2^{\,n}/(n\chi^2)\right)$, which grows exponentially in $n$. This is a
dimension-counting (local, noiseless identifiability) statement, not a sample-complexity or
noise-robustness theorem; it quantifies why a prior that restricts inference to $\mathcal M$ can succeed
in the few-measurement regime where unstructured estimators cannot.
\end{proposition}

\paragraph{Approach~A recovery theory (deferred).} The generative-prior route admits a prior-restricted stable-recovery bound under a manifold restricted-isometry \emph{hypothesis} (Prop.~\ref{prop:stable}) and a best-of-$N$ selection guarantee (Prop.~\ref{prop:bestof}); because the RIP is assumed rather than established for a concrete Pauli design, we state both in App.~\ref{app:Atheory} to keep the main-text theory proportionate to what we prove.

\begin{proposition}[Distribution-free coverage of recalibrated conformal sets]
\label{prop:cov}
Let calibration states $\{\ket{\psi_i}\}_{i\in\mathcal C}$ and a test state $\ket{\psi_\star}$ be
exchangeable draws from the family. For an observable $O$ let $\mu(\cdot),\sigma(\cdot)$ be the
predictive mean and scale from the estimator's dropout ensemble and $r(\psi)=|O(\psi)-\mu(\psi)|/\sigma(\psi)$ the
nonconformity score, and let $Q$ be the $\lceil(1-\alpha)(|\mathcal C|+1)\rceil$-th smallest
calibration score. Then the recalibrated interval
$\mathcal I(\psi_\star)=[\mu(\psi_\star)\pm Q\,\sigma(\psi_\star)]$ satisfies
$\;\Pr\!\big(O(\psi_\star)\in\mathcal I(\psi_\star)\big)\ge 1-\alpha$, with no assumption on the prior
or the sampler.
\end{proposition}

\paragraph{Scope of the theory (a formalization).} We position the theory as scaffolding, not a main contribution. Prop.~\ref{prop:adv} is a degrees-of-freedom/local-identifiability count, not a sample-complexity theorem; Prop.~\ref{prop:localdesign} exhibits a \emph{constructive} $O(n)$ local spanning design; Prop.~\ref{prop:coorddep} formalizes that a prior over cores is coordinate-dependent; and Prop.~\ref{prop:cov} is standard split-conformal, which is what our calibrated intervals rely on. What we do \emph{not} prove: how many measurements a given design needs as a function of the manifold's covering number, curvature, and bond dimension, and quantitative error bounds for the amortized conditional map---both are open.

\section{Experiments}
\label{sec:exp}
\paragraph{Setup.} We evaluate on two $n{=}6$, $\chi{=}2$ families. The \emph{generic} random-MPS family
draws fixed random base cores plus i.i.d.\ complex-Gaussian perturbations ($\sigma{=}0.35$), canonicalized;
it is moderately diverse (nearest-training-neighbour fidelity $\approx0.82$). The \emph{physical}
shallow-circuit family is an $R_y(\theta_1)$--CNOT-ladder--$R_y(\theta_2)$ circuit with random angles
($\theta_1\!\sim\!\tfrac\pi2+0.6\mathcal N,\ \theta_2\!\sim\!0.6\mathcal N$), which is exactly a $\chi{=}2$
MPS and, at this spread, genuinely diverse (medoid fidelity to test states only $0.37$). Measurements are
noisy Pauli expectations $o_j=\langle P_j\rangle+\varepsilon_j$ with shot noise of variance
$(1{-}\langle P_j\rangle^2)/S$, $S{=}200$. Unless stated we use the informative local design $\mathcal D$
(all single-qubit $X_i,Y_i,Z_i$ and nearest-neighbour $P_iP_{i+1}$; $63$ observables at $n{=}6$). The
conditional set-encoder is trained once per family ($2\!\times\!10^4$/$6\!\times\!10^3$ states, $30$k AdamW
steps); all fidelities are on $120$--$200$ held-out states. Code and hyperparameters are in
App.~\ref{app:details}.

\begin{figure}[t]
\centering
\includegraphics[width=0.95\textwidth]{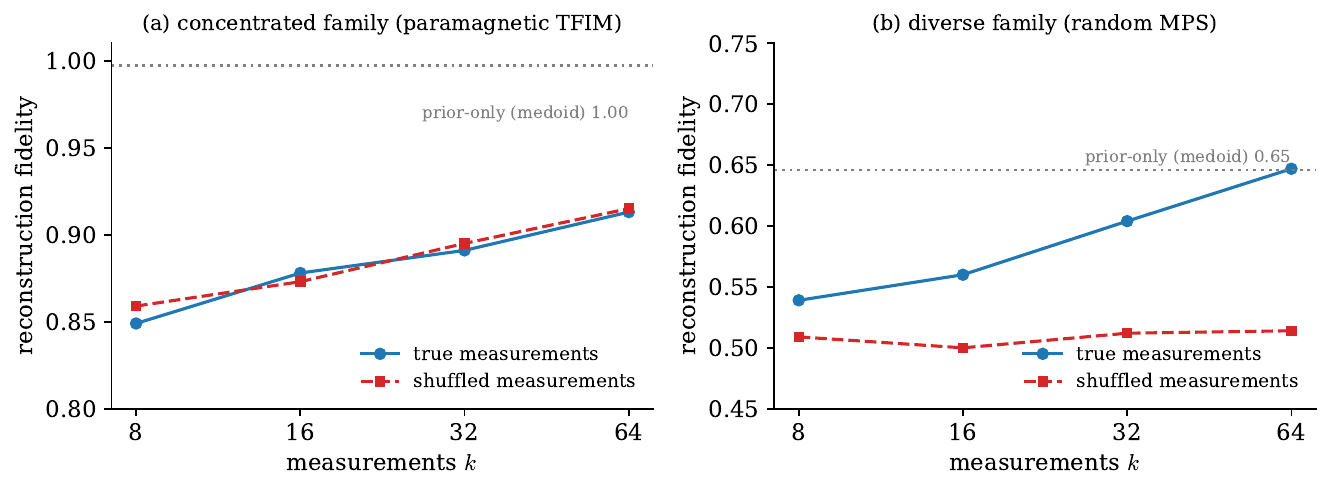}
\caption{Integrity controls. \textbf{(a)} On a concentrated family, a $k{=}0$ prior-only medoid ($0.997$)
dominates and true$\,\approx\,$shuffled: the measurements are not used, so ``few-measurement fidelity'' is
memorization. \textbf{(b)} On a diverse family the medoid is far from optimal and true reconstruction
separates from shuffled, with the gap growing in the budget $k$---genuine measurement information. Both
controls are prerequisites for interpreting any learned-QST fidelity.}
\label{fig:control}
\end{figure}

\paragraph{Integrity controls: beat prior-only, and use the measurements.}
We first run the two controls that, we argue, any learned-QST method should report. \emph{Prior-only
($k{=}0$):} predict each test state by a fixed central estimate that uses no measurements---the family
medoid (the bank state maximizing mean fidelity to the family) or the normalized mean bank state.
\emph{Shuffled:} reconstruct from a \emph{different} state's measurements (a random permutation of the test
batch); if this matches the true-measurement reconstruction, the measurements are unused and the method is
doing nearest-neighbour prediction, not tomography. Fig.~\ref{fig:control} shows why these matter. On a
\emph{concentrated} family---disordered paramagnetic TFIM ground states, all close to
$\ket{+}^{\otimes n}$---the medoid already reaches fidelity $0.997$ with \emph{zero} measurements, and true
and shuffled reconstructions coincide to $\pm0.01$ at every budget: there is nothing to reconstruct, and a
naive ``high fidelity at few measurements'' is entirely the prior. On a genuinely \emph{diverse} family the
medoid is far from optimal ($0.65$), true-measurement reconstruction clearly separates from shuffled, and
the separation grows with the budget---the signature of real measurement information. We therefore report
on diverse families, where passing the controls is meaningful.

\paragraph{Approach~A: a gold-standard-validated posterior that leans on the prior.}
Before conditioning, we evaluate the generative-prior route (\S~Method). Its output is a genuine posterior:
against a self-normalized importance sampler drawn from the learned prior and reweighted by the exact
likelihood (the gold standard), the sampler's posterior means and per-observable marginals agree closely
($n{=}6$: posterior-mean correlation $0.98$, median $1$D energy distance $0.033$, ESS $\sim\!200/10^5$), so
it targets the intended posterior rather than a relabelled point estimate. Approach~A also clearly beats
classical estimators at few measurements---maximum-likelihood, a generic-prior pCN sampler, and
classical-shadow reconstruction all sit far below it on structured families---and it degrades gracefully as
states leave the family rather than hallucinating. \emph{Yet the integrity controls temper the
interpretation.} On the concentrated family a $k{=}0$ prior-only estimate already matches it and shuffling
the measurements barely changes the reconstruction (Fig.~\ref{fig:control}a); even on the diverse random
family the prior-only medoid is competitive with the flow reconstruction at the budgets tested. Approach~A
is thus a valid, calibrated Bayesian posterior whose \emph{point accuracy} at few measurements is largely
inherited from the prior---exactly the gap that motivates conditioning directly on the measurements
(Approach~B), to which we now turn.

\begin{figure}[t]
\centering
\includegraphics[width=0.92\textwidth]{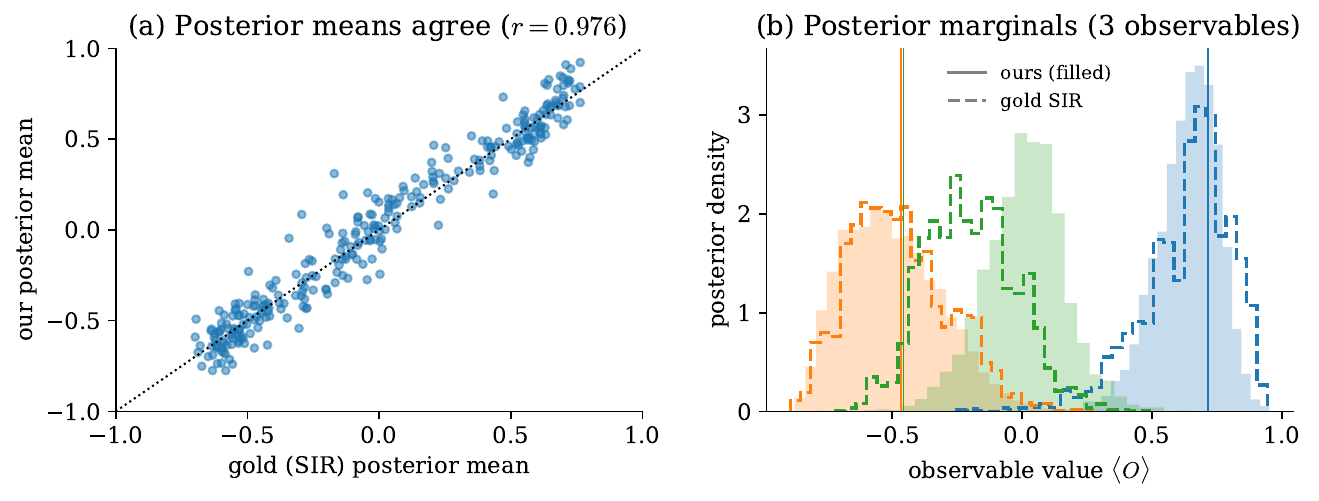}
\caption{Approach~A targets the intended posterior ($n{=}6$). \textbf{(a)} Posterior means of the flow
sampler vs.\ a gold-standard SIR posterior, over $25$ states and $2n$ local observables, lie on the
diagonal ($r{=}0.98$). \textbf{(b)} Example posterior marginals (three observables): the sampler's draws
(filled) overlap the SIR posterior (dashed); vertical lines are the true values. This confirms the
generative-prior route is a genuine posterior, not a relabelled point estimate.}
\label{fig:postval}
\end{figure}

\begin{figure}[t]
\centering
\includegraphics[width=0.9\textwidth]{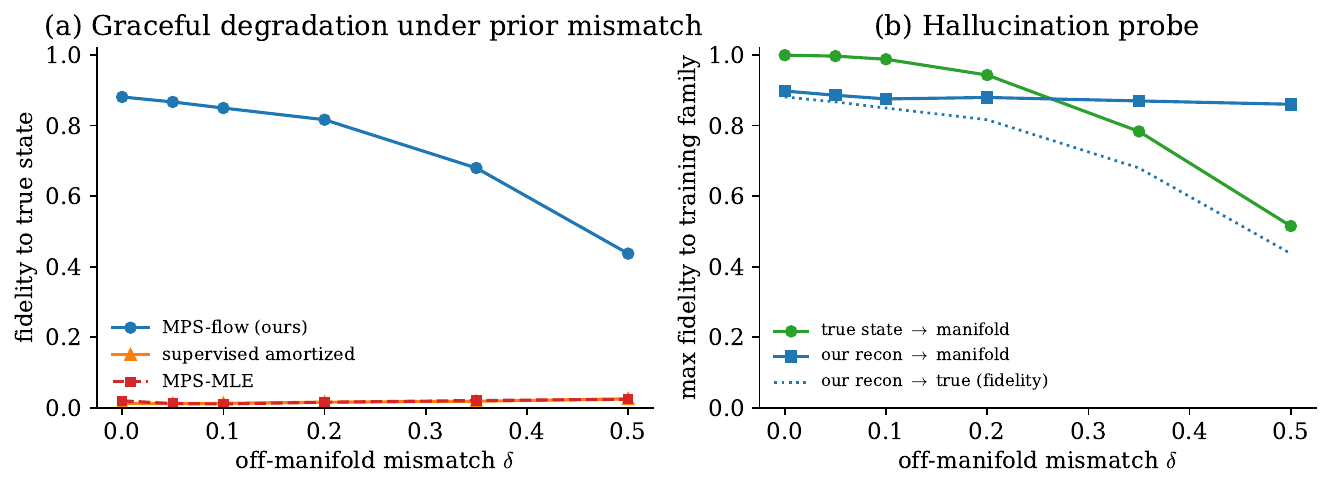}
\caption{Out-of-distribution behaviour of Approach~A (physical family, $n{=}6$). \textbf{(a)} As the true
state is pushed off the manifold (mismatch $\delta$), fidelity degrades gracefully. \textbf{(b)}
Hallucination probe: the reconstruction stays near the training family (blue) while the true state leaves
it (green); the reconstruction's fidelity to the \emph{true} state (dotted) declines with $\delta$ rather
than staying spuriously high.}
\label{fig:ood}
\end{figure}

\paragraph{Approach~A is robust out-of-distribution.}
Because Approach~A is a prior-based method, we stress-test it off the training manifold by mixing in a
Haar-random direction, $\ket{\psi_\delta}\propto(1{-}\delta)\ket{\psi_{\mathrm{fam}}}+\delta\ket{\psi_{\mathrm{Haar}}}$
(Fig.~\ref{fig:ood}). Reconstruction fidelity degrades \emph{gracefully and monotonically}
($0.88\!\to\!0.44$ as $\delta$ grows) rather than collapsing at a threshold, and a hallucination probe
shows the reconstruction is pulled toward the training family but its fidelity to the true state declines
in step with $\delta$ rather than staying spuriously high. This graceful degradation, and the calibrated
intervals, are genuine strengths of the generative-prior route---orthogonal to, and retained alongside,
the higher point fidelity of Approach~B.

\begin{figure}[t]
\centering
\begin{minipage}[t]{0.34\textwidth}\centering\includegraphics[width=\textwidth]{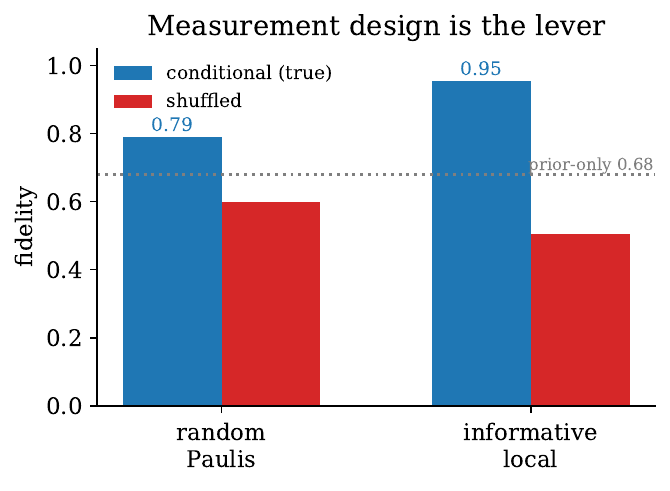}\end{minipage}\hfill
\begin{minipage}[t]{0.36\textwidth}\centering\includegraphics[width=\textwidth]{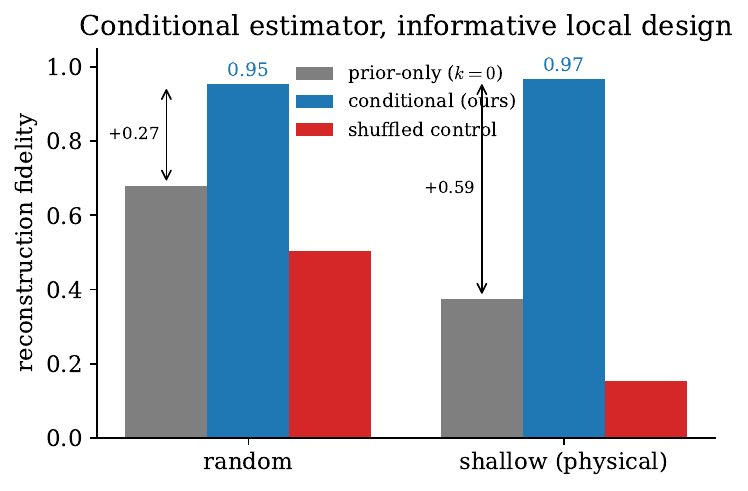}\end{minipage}\hfill
\begin{minipage}[t]{0.28\textwidth}\centering\includegraphics[width=\textwidth]{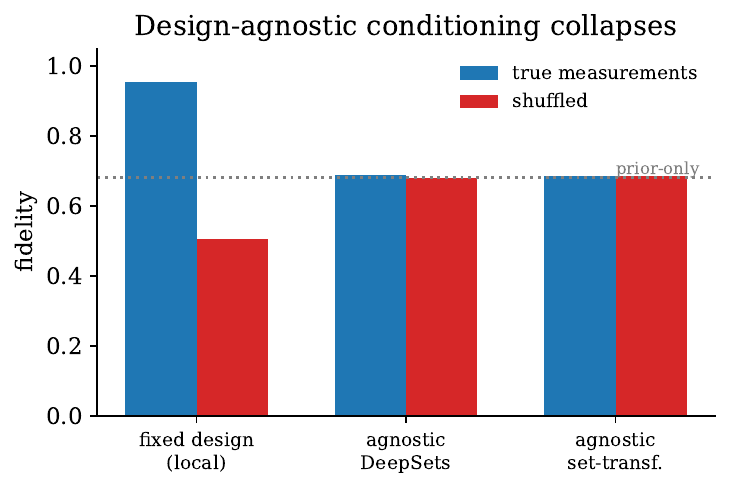}\end{minipage}
\caption{\textbf{Left:} the measurement design is the lever---informative local Paulis lift fidelity from
$0.79$ to $0.95$ over random Paulis (random family). \textbf{Middle:} headline result---the conditional
estimator beats the $k{=}0$ prior-only baseline (arrows: $+0.27$ random, $+0.59$ shallow) and the shuffled
control collapses. \textbf{Right:} making the estimator design-\emph{agnostic} (training over variable
random subsets) collapses it to the family mean (true $=$ shuffled), for both a DeepSets and a
set-transformer encoder.}
\label{fig:design}
\end{figure}

\paragraph{The measurement design is the lever.}
Conditioning on measurements is necessary but not sufficient: with \emph{random} Pauli strings the
conditional model beats prior-only only modestly on the random family ($0.79$ vs.\ $0.67$, gain $+0.12$;
true$-$shuffled $+0.19$), because a random weight-$w$ Pauli has expectation exponentially small in $w$ and
carries little information about a low-entanglement state. Switching to the theory-motivated
\emph{informative local} design (Prop.~\ref{prop:localdesign}) is decisive (Fig.~\ref{fig:design},
Table~\ref{tab:main}): fidelity rises to $0.954{\pm}0.021$ (random) and $0.967{\pm}0.046$ (shallow), the
prior-only gain to $+0.27$ and $+0.59$, and the shuffled control collapses the reconstruction to $0.51$
and $0.15$ (true$-$shuffled up to $+0.81$---feeding another state's local measurements yields a wrong
state). The
design change, not the encoder or more training, is what turns a modest, memorization-prone estimator into
a high-fidelity one.

\begin{table}[t]
\centering
\caption{Conditional estimator, $n{=}6$, over $120$--$200$ held-out states (conditional fidelity reported
as mean$\pm$std across states; per-state $q_{10}$--$q_{90}$ is $0.924$--$0.974$ for random and
$0.927$--$0.992$ for shallow---the spread is the dispersion of reconstruction quality across states, and
is deterministic across training seeds). Point fidelity vs.\ the $k{=}0$ prior-only baseline
(medoid/mean); the shuffled control (reconstruction from another state's measurements) collapses the
fidelity, confirming the measurements are used; conformal recalibration attains near-nominal $90\%$
coverage with sharp intervals. The last row shows the random-Pauli design: switching random$\to$local is
what makes conditioning work.}
\label{tab:main}
\resizebox{\textwidth}{!}{%
\begin{tabular}{llcccccc}
\toprule
family & design & prior-only & \textbf{conditional} & gain & shuffled & true$-$shuf. & coverage / width \\
\midrule
random  & local  & 0.680 & $\mathbf{0.954{\pm}0.021}$ & $+0.27$ & 0.505 & $+0.45$ & $89.5\%$ / $0.20$ \\
shallow (phys.) & local & 0.375 & $\mathbf{0.967{\pm}0.046}$ & $+0.59$ & 0.154 & $+0.81$ & $89.4\%$ / $0.23$ \\
\midrule
random  & random Paulis & 0.680 & 0.79 & $+0.11$ & 0.60 & $+0.19$ & --- \\
\bottomrule
\end{tabular}}
\end{table}

\begin{figure}[t]
\centering
\includegraphics[width=0.88\textwidth]{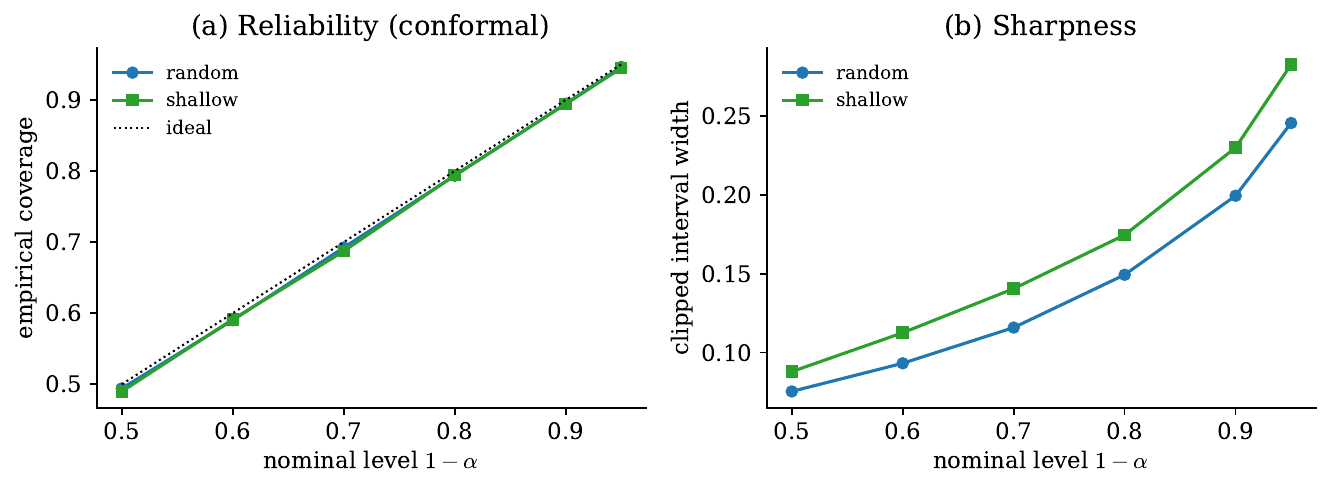}
\caption{Calibration of the dropout predictive ensemble over the $2n$ local $Z/X$ observables (conformal, $E{=}24$
samples, $10$ splits). \textbf{(a)} Empirical coverage tracks the nominal level across the range (ideal:
diagonal). \textbf{(b)} Interval width stays small (physical range $[-1,1]$), because the high-fidelity
reconstruction concentrates the predictive ensemble. Raw (uncalibrated) coverage is $69$--$72\%$ at the $90\%$ level.}
\label{fig:calib}
\end{figure}

\paragraph{Calibrated, sharp uncertainty.}
Beyond a point estimate we return a \emph{predictive ensemble} (not a validated Bayesian posterior---that
is Approach~A): keeping dropout active at test time, $E{=}24$ stochastic passes give a predictive spread
and hence, for each of the $2n$ local $Z/X$ observables, an ensemble mean $\mu_O$ and scale $\sigma_O$. The raw intervals are over-confident (empirical coverage $69$--$72\%$ of a
nominal $90\%$ interval; Fig.~\ref{fig:calib}a, dashed), reflecting that MC-dropout under-estimates
uncertainty. Split-conformal recalibration restores coverage to $89.5\pm1.6\%$ (random) and $89.4\pm1.5\%$
(shallow) over $10$ random calibration/test splits (Table~\ref{tab:calib}), with the reliability curve
tracking the diagonal across nominal levels (Fig.~\ref{fig:calib}a). Crucially the intervals are also
\emph{sharp}---clipped physical widths of only $0.20$ and $0.23$ (Fig.~\ref{fig:calib}b)---because the
high-fidelity reconstruction concentrates the predictive ensemble. A prior-only estimate provides no calibrated
uncertainty at all; the frequentist coverage guarantee here is a property the point-estimate baselines
cannot offer.

\paragraph{Calibrated predictions for \emph{unmeasured} observables.} Intervals on the measured $Z/X$
observables could in principle be built directly from their own shot counts, so the sharper test of
predictive value is on observables the protocol \emph{never measured}. We take $18$ held-out
observables---all long-range ($\ge2$-apart) $Z_iZ_j$ correlators and the weight-$3$ $Z_iZ_{i{+}1}Z_{i{+}2}$
and $X_iX_{i{+}1}X_{i{+}2}$ strings, none of which are in the \texttt{local} design---and predict them from
the reconstruction with the same MC-dropout$+$conformal pipeline. The reconstruction predicts these
never-measured observables with mean absolute error $0.043$ (random) / $0.053$ (shallow), versus $0.133$ /
$0.267$ for a prior-only estimate ($3$--$5\times$ better), and the conformalized intervals attain
$89.7\pm0.8\%$ / $89.6\pm1.8\%$ coverage at width $\approx0.20$. A shot-based interval for these observables
\emph{does not exist}---they were never measured---so this is genuine predictive value from the
reconstruction, not a repackaging of the input shots, and it is what distinguishes our calibrated
uncertainty from a trivial per-observable binomial interval on the inputs.

\begin{table}[t]
\centering
\caption{Conformal calibration of the local-$Z/X$ observable intervals (nominal $90\%$, mean$\pm$std over
$10$ splits). Raw MC-dropout is over-confident; conformal recalibration restores coverage, and the
intervals are sharp because reconstruction fidelity is high.}
\label{tab:calib}
\begin{tabular}{lccc}
\toprule
family & raw coverage & \textbf{conformal coverage} & clipped width \\
\midrule
random & $68.9\%$ & $\mathbf{89.5\pm1.6\%}$ & $0.199\pm0.005$ \\
shallow (physical) & $72.2\%$ & $\mathbf{89.4\pm1.5\%}$ & $0.230\pm0.005$ \\
\bottomrule
\end{tabular}
\end{table}

\paragraph{Design-agnostic conditioning collapses (an honest limitation).}
The estimator above is tied to the fixed design it trains on. We tried to make it design-\emph{agnostic} by
training over variable random Pauli subsets and testing on unseen designs. Neither a DeepSets encoder nor a
$4$-layer set-transformer with attention pooling learns to exploit arbitrary designs: both collapse to the
family mean, with true $=$ shuffled $=0.69\approx$ prior-only and gain $\approx0$ (Fig.~\ref{fig:design},
right). This is a genuine negative result---amortizing over arbitrary measurement designs, or over a
parameterized family of designs, is open---and it sharpens the scope: our claims are for a \emph{fixed
informative protocol}, which is the standard setting in practice.

\begin{figure}[t]
\centering
\includegraphics[width=0.9\textwidth]{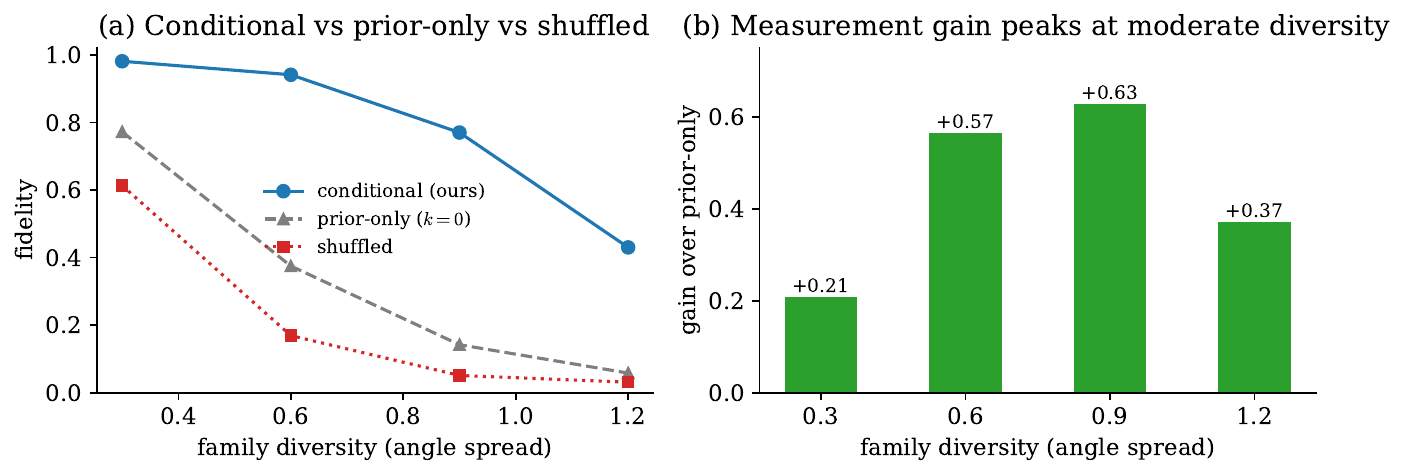}
\caption{Family-diversity sweep (shallow-circuit family, local design). \textbf{(a)} As diversity grows the
$k{=}0$ prior-only fidelity falls and conditional reconstruction eventually gets harder, but the shuffled
control stays low throughout. \textbf{(b)} The \emph{measurement gain} over prior-only peaks at moderate
diversity: too concentrated and the prior alone suffices; too diverse and reconstruction is hard.}
\label{fig:diversity}
\end{figure}

\paragraph{Ablations: design richness, encoder, diversity, shots.}
We isolate what drives the result (Table~\ref{tab:abl}, Fig.~\ref{fig:diversity}). \emph{(i) Design
richness.} Single-qubit Paulis alone (\texttt{local1}, $18$ obs) already beat prior-only ($0.910$ random)
but underperform on the physical family ($0.666$ shallow), where the nearest-neighbour weight-$2$ terms
matter; the default \texttt{local} set ($63$) reaches $0.954$/$0.967$; the theory-complete $3$-site-window
design (\texttt{local3}, $207$ obs; Prop.~\ref{prop:localdesign}) is comparable ($0.950$/$0.957$)---so the
cheap $2$-local subset is empirically sufficient and the theory-complete superset is not needed at this
budget. In all cases the shuffled control collapses, confirming the measurements are used. \emph{(ii)
Encoder.} For a \emph{fixed} design, a plain MLP on the fixed-order value vector (ignoring the Pauli
identities and set structure) matches or slightly exceeds the DeepSets set-encoder ($0.967$/$0.991$ vs
$0.954$/$0.967$): with a fixed protocol the performance comes from the MPS parameterization $+$ local
design $+$ fidelity loss, not from permutation-invariance. The set structure is only motivated by the
design-\emph{agnostic} goal, where it (and a set-transformer) nonetheless collapse. \emph{(iii)
Known-ansatz baseline.} A reviewer-style stronger baseline exploits the fact that the shallow family is
generated by $2n$ circuit angles: we regress those angles directly from the $63$ local observables and
rebuild the state. Despite using knowledge of the exact ansatz that our method never assumes, it reaches
only $0.942{\pm}0.089$---\emph{below} our $0.967{\pm}0.046$ and far more variable. To rule out that this is
merely an unfavorable angle-MSE objective, we also train the \emph{same} circuit-angle decoder with the
identical gauge-invariant \emph{state-fidelity} loss our method uses (predicting angles, rebuilding the
state, minimizing $1{-}|\langle\psi_\star|\psi(\hat\theta)\rangle|^2$); it improves only to $0.947{\pm}0.098$,
still below the MPS estimator, because the angle representation is many-to-one and non-convex whereas the
canonical-core target is better conditioned. So even a structure-aware regressor, given our exact objective,
does not dominate the ansatz-agnostic MPS estimator. \emph{(iv) Diversity.}
Sweeping the shallow family's angle spread (Fig.~\ref{fig:diversity}) traces the expected behaviour: on
concentrated families prior-only is already high and the measurement gain is small; the gain peaks at
moderate diversity ($+0.63$); on very diverse families reconstruction is harder but still clearly beats
prior-only. \emph{(v) Shots.} Reducing $S$ from $800$ to $50$ shots per setting lowers fidelity gracefully
(App.~\ref{app:details}), as expected from the $1/\sqrt S$ measurement noise.

\begin{table}[t]
\centering
\caption{Ablations (conditional fidelity, $n{=}6$, local design unless noted). \emph{Design richness:}
the cheap $2$-local set (\texttt{local}) is sufficient; single-qubit-only underperforms on the physical
family, and the theory-complete $3$-site design does not improve over \texttt{local}. \emph{Encoder:} for a
fixed design a plain MLP matches the set-encoder, so permutation-invariance is not what drives performance.
Shuffled fidelity in parentheses.}
\label{tab:abl}
\begin{tabular}{lcc}
\toprule
setting & random family & shallow (physical) \\
\midrule
\texttt{local1} (single-qubit, $18$) & $0.910$ ($0.53$) & $0.666$ ($0.19$) \\
\texttt{local} (default, $63$) & $\mathbf{0.954}$ ($0.51$) & $\mathbf{0.967}$ ($0.15$) \\
\texttt{local3} (theory-complete, $207$) & $0.950$ ($0.50$) & $0.957$ ($0.16$) \\
\midrule
encoder: DeepSets (\texttt{local}) & $0.954$ & $0.967$ \\
encoder: plain MLP (\texttt{local}) & $0.967$ & $0.991$ \\
\bottomrule
\end{tabular}
\end{table}

\paragraph{Head-to-head with classical local reconstruction, and what the learned method actually buys.}
The most informative baseline is the classical one our design is motivated by: a \emph{prior-free} MPS
maximum-likelihood fit to the \emph{same} local Pauli values (a stand-in for Cramer-style local
reconstruction), run per state with random restarts. We compare it to our amortized estimator across the
design ladder (Table~\ref{tab:mle}, shallow family). Two honest conclusions follow. First, the
\emph{measurement design is the dominant lever for both methods}: classical MPS-MLE climbs from $0.633$
(\texttt{local1}, under-determined) to $0.953$ (\texttt{local}) to $0.989$ (\texttt{local3}, theory-complete),
mirroring the learned estimator---so the informative local design, not the learned prior per se, is what
makes few-measurement reconstruction possible. Second, the learned prior's \emph{advantage is confined to
the under-determined regime}: at \texttt{local1} it edges out MLE ($0.666$ vs.\ $0.633$), at \texttt{local}
it is a marginal $+0.014$, and at the theory-complete \texttt{local3} classical MLE is actually
\emph{better} ($0.989$ vs.\ $0.957$)---with complete data the prior is unnecessary and our fixed-capacity
network is the bottleneck. We therefore do \emph{not} claim the learned estimator dominates classical
reconstruction on fidelity. What it buys is \emph{amortization}---a single forward pass instead of a
per-state nonconvex optimization---and \emph{calibrated predictive uncertainty} (including for unmeasured
observables), neither of which the per-state MLE provides. The earlier near-zero MLE numbers
(Table~\ref{tab:Abase}) are for \emph{random} Pauli designs; on the informative-local design MLE is strong,
which is the same ``design is the lever'' point.

\begin{table}[t]
\centering
\caption{Classical prior-free MPS-MLE vs.\ our amortized estimator across the design ladder (shallow family,
$n{=}6$, $S{=}200$). The design is the lever for both; the learned prior helps only where the design is
under-determined (\texttt{local1}), is marginal on \texttt{local}, and is unnecessary at the theory-complete
\texttt{local3}. The learned method's value is amortization $+$ calibrated UQ, not a fidelity margin.}
\label{tab:mle}
\begin{tabular}{lccc}
\toprule
design & \texttt{local1} ($18$) & \texttt{local} ($63$) & \texttt{local3} ($207$) \\
\midrule
classical MPS-MLE (per-state, prior-free) & $0.633{\pm}0.220$ & $0.953{\pm}0.017$ & $\mathbf{0.989{\pm}0.004}$ \\
amortized estimator (ours) & $\mathbf{0.666}$ & $\mathbf{0.967}$ & $0.957$ \\
\bottomrule
\end{tabular}
\end{table}

\paragraph{Family-aware simple baselines, and why the MPS core.} Two further baselines isolate what the
representation contributes. (i) A \emph{measurement-space $k$-NN}: for a test measurement vector, return the
state of the nearest (or mean of the $5$ nearest) training states in measurement space. This trivial
memorization reaches $0.80$/$0.82$ ($1$-NN) and $0.90$/$0.86$ ($5$-NN) on shallow/random---useful, but
clearly below the amortized estimator ($0.967$/$0.954$), so the method is not merely retrieving a nearby
training state. (ii) A \emph{direct dense-state MLP} that outputs the $2^n$ amplitudes instead of MPS cores,
same loss and protocol, reaches $0.979$/$0.958$---i.e.\ at $n{=}6$ it \emph{matches or slightly exceeds} the
MPS-core estimator. We report this openly: the MPS-core parameterization is \emph{not} what buys accuracy at
small $n$. Its value is scalability---the dense-state output has $2^n$ parameters and is infeasible at the
$n{=}16,20$ where the MPS core (with $O(n\chi^2)$ parameters and native $O(n\chi^3)$ contraction) still
runs. The core representation is a scalability choice, not a small-$n$ accuracy claim.

\paragraph{Higher bond dimension.}
All results so far use the $\chi{=}2$ physical family; we test more entangled targets by sweeping the bond
dimension of the random-MPS family at $n{=}6$ with the same \texttt{local} design (Table~\ref{tab:chi}). As
$\chi$ grows the states are harder and, notably, the weight-$2$ \texttt{local} design is \emph{theoretically
sub-complete} for $\chi{>}2$ (Prop.~\ref{prop:localdesign} asks for $\ell{=}5$-site windows at $\chi{=}4$,
not $\ell{=}3$). Reconstruction fidelity nonetheless degrades only gently ($0.954{\to}0.908{\to}0.878$ for
$\chi{=}2,3,4$) and the estimator keeps beating prior-only by $+0.30$ and passing the shuffled control
(true$-$shuffled $\ge+0.45$): the learned prior compensates for the sub-complete design, and the method does
not break at $\chi{=}4$. Closing the theory--practice gap with the $\ell{=}5$ design at higher $\chi$ is
natural future work. As a further check that we are not merely fitting the specific one-layer
$R_y$--CNOT--$R_y$ ansatz, we also train on a \emph{deeper} two-layer brickwork family ($R_y$--CNOT--$R_y$--CNOT--$R_y$,
entanglement up to $\chi{=}4$): reconstruction reaches $0.782\pm0.140$, still well above the $k{=}0$
prior-only baseline ($+0.56$) and decisively passing the shuffled control (true$-$shuffled $+0.70$)---the
method degrades gracefully with circuit depth without collapsing.

\begin{table}[t]
\centering
\caption{Bond-dimension sweep (random-MPS family, $n{=}6$, \texttt{local} design). Fidelity degrades gently
and the measurement gain persists even though \texttt{local} is theoretically sub-complete for $\chi{>}2$.}
\label{tab:chi}
\begin{tabular}{lccc}
\toprule
bond dimension $\chi$ & $2$ & $3$ & $4$ \\
\midrule
conditional fidelity   & $0.954{\pm}0.021$ & $0.908{\pm}0.036$ & $0.878{\pm}0.063$ \\
prior-only ($k{=}0$)   & $0.680$ & $0.594$ & $0.581$ \\
gain over prior-only   & $+0.27$ & $+0.31$ & $+0.30$ \\
true $-$ shuffled      & $+0.45$ & $+0.50$ & $+0.45$ \\
\bottomrule
\end{tabular}
\end{table}

\begin{figure}[t]
\centering
\includegraphics[width=0.92\textwidth]{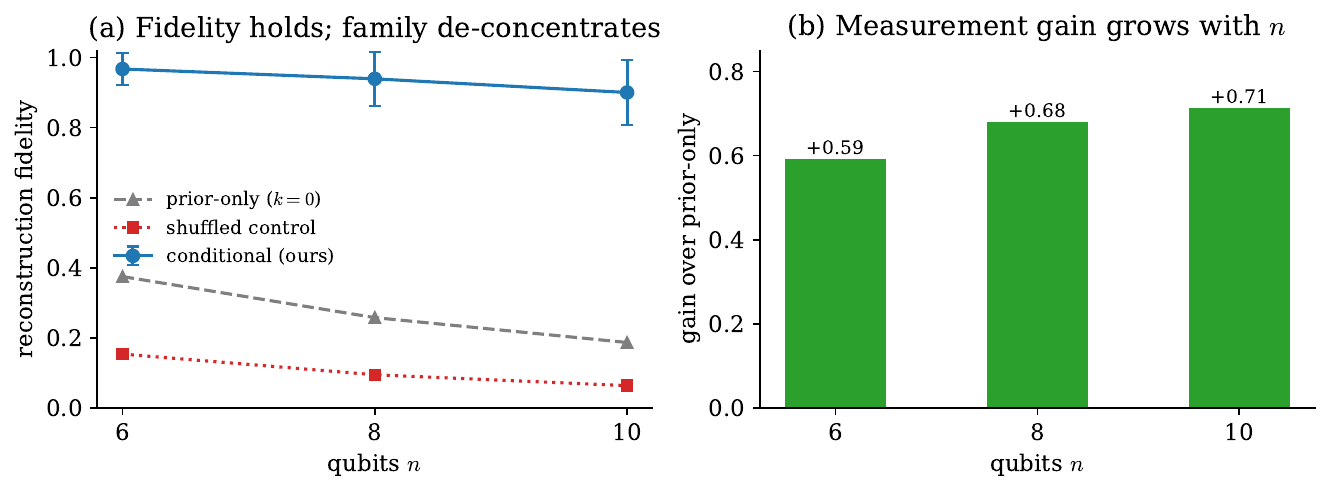}
\caption{Conditional reconstruction vs.\ system size (shallow family, informative local design, $30$k-step
budget). \textbf{(a)} Fidelity (mean$\pm$std across states) declines only gently while the $k{=}0$
prior-only fidelity falls steeply and the shuffled control stays low. \textbf{(b)} The measurement gain
over prior-only \emph{grows} with $n$---at larger sizes the reconstruction relies more, not less, on the
measurements.}
\label{fig:scalingcond}
\end{figure}

\paragraph{Reconstruction quality scales with system size.}
Beyond the computational-feasibility demonstration below, we test whether \emph{tomography performance}
holds as $n$ grows, by running the full conditional pipeline on the shallow family with the informative
local design at $n{=}6,8,10$ (same $30$k-step budget; Fig.~\ref{fig:scalingcond}). Reconstruction fidelity
declines only gently ($0.967{\to}0.939{\to}0.900$), \emph{not} collapsing. More telling for a tomography
claim: as $n$ grows the family de-concentrates, so the $k{=}0$ prior-only fidelity falls steeply
($0.375{\to}0.258{\to}0.187$) while the shuffled control stays low ($0.154{\to}0.095{\to}0.064$)---so the
\emph{measurement gain} over prior-only actually \emph{grows} with $n$ ($+0.59{\to}+0.68{\to}+0.71$) and
the true$-$shuffled gap stays $\approx0.84$. The reconstruction is thus increasingly driven by genuine
measurement information at larger $n$, not by the prior; this is the sense in which performance, and not
just the implementation, scales. (Dense-statevector training limits this sweep to $n{\le}10$; the native
contraction below removes the $2^n$ memory cost but we do not yet pair it with high-fidelity training at
$n{>}10$.)

\begin{figure}[t]
\centering
\includegraphics[width=0.6\textwidth]{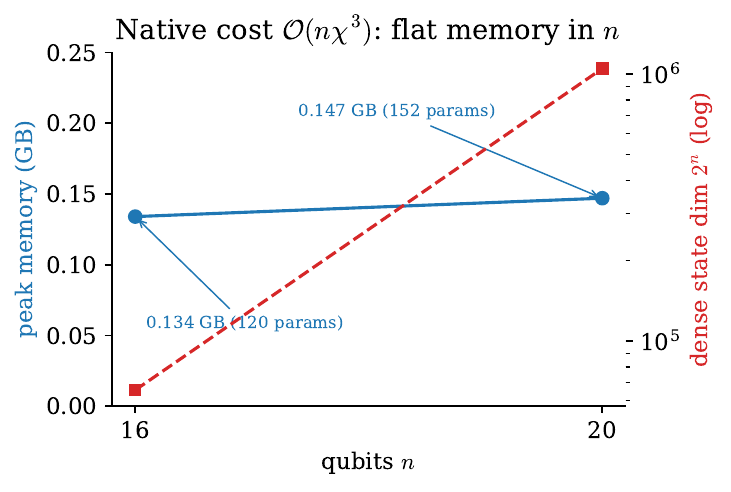}
\caption{Native $O(n\chi^3)$ contraction: peak GPU memory is flat in $n$ ($0.134$~GB at $n{=}16$,
$0.147$~GB at $n{=}20$; $120$/$152$ complex core parameters) while the dense state dimension $2^n$ grows
exponentially ($65{,}536\!\to\!1{,}048{,}576$). Running the full conditional pipeline; a
computational-feasibility result, not high-fidelity tomography at these sizes.}
\label{fig:scale}
\end{figure}

\paragraph{Scalability: native tensor-network contraction to $n{=}20$.}
The core parameterization is polynomial in $n$ ($O(n\chi^2)$ parameters); to show the \emph{implementation}
is too, we compute every Pauli expectation $\bra{\psi}\!\bigotimes_i P_i\ket{\psi}$ and overlap by
left-to-right transfer-matrix contraction at $O(n\chi^3)$, never materializing the $2^n$ state vector (the
primitives match the dense contraction to $\le\!10^{-8}$; App.~\ref{app:details}). Running the full
conditional pipeline on the random-MPS family, peak GPU memory is $0.134$~GB at $n{=}16$ and $0.147$~GB at
$n{=}20$---essentially \emph{flat} in $n$ while the dense dimension $2^n$ grows from $65{,}536$ to
$1{,}048{,}576$, using only $120$ and $152$ complex core parameters (Fig.~\ref{fig:scale}). This is a
computational-scalability demonstration: fidelity at these sizes is low
(the harder large-$n$ family, reduced budget), not high-fidelity tomography at $20$ qubits.

\begin{table}[t]
\centering
\caption{Closed-loop on IBM \texttt{ibm\_aachen} (Heron): the \emph{local} design ($63$ Paulis) measured on
$5$ held-out states in one batched \texttt{EstimatorV2} job, then reconstructed by the conditional
estimator. Reconstruction from hardware nearly matches reconstruction from noiseless values, and both far
exceed the $k{=}0$ prior-only baseline.}
\label{tab:hw}
\resizebox{\textwidth}{!}{%
\begin{tabular}{lccccc}
\toprule
device $|hw{-}\text{ideal}|$ & convention check & \textbf{fid.\ from HW} & fid.\ from ideal & prior-only ($k{=}0$) & per-state HW range \\
\midrule
$0.024$ & $<\!10^{-7}$ & $\mathbf{0.971{\pm}0.006}$ & $0.977$ & $0.283$ & $0.964$--$0.979$ \\
\bottomrule
\end{tabular}}
\end{table}

\paragraph{Closed-loop reconstruction on IBM hardware.}
We close the loop end-to-end on real hardware with the \emph{same} informative-local design as the method.
On \texttt{ibm\_aachen} ($156$-qubit Heron) we prepare $5$ held-out shallow-circuit states and measure the
$63$ local Paulis with \texttt{EstimatorV2} in a single batched job (device error mean
$|hw{-}\text{ideal}|=0.024$; Qiskit-vs-torch conventions verified to $<\!10^{-7}$; Table~\ref{tab:hw}). We
then feed these \emph{hardware-measured} values to the conditional estimator: it reconstructs the $5$ states
at fidelity $\mathbf{0.971{\pm}0.006}$ (per-state $0.964$--$0.979$). This is genuine tomography, not
memorization: the $k{=}0$ prior-only medoid reaches only $0.283$ on these states, so the reconstruction is
$+0.69$ from using the measurements. And it is robust to device noise---reconstructing from the
\emph{noiseless} ideal values gives $0.977$, so the $0.024$ device error costs only $0.006$ in fidelity.
This is the reviewer-requested closed loop (hardware measurements $\to$ estimator $\to$ reconstruction
fidelity), not merely a measurement-pipeline demonstration.

\section{Discussion: when each approach helps}
\label{sec:discussion}
The two approaches are complementary, and the integrity controls make the trade-off precise. Approach~A---%
the generative prior with measurement-guided posterior inference---is a genuine, gold-standard-validated
Bayesian posterior: it provides full posterior samples, calibrated intervals, graceful out-of-distribution
degradation, and it dominates classical few-measurement estimators that lack the family prior
(Table~\ref{tab:Abase}). Its weakness, exposed by the $k{=}0$ prior-only control, is that on concentrated
families its \emph{point} accuracy is largely inherited from the prior and the measurements are
under-used---especially with the random Pauli designs typical of prior-based inverse-problem solvers.
Approach~B---the amortized measurement-conditioned estimator---directly extracts measurement information
and, with the informative-local design, reconstructs at markedly higher fidelity while clearly beating
prior-only and passing the shuffled control; its cost is that it is a point estimator with heuristic
(dropout$+$conformal) uncertainty rather than a validated posterior, and is tied to a fixed measurement
design. A practitioner wanting a calibrated posterior and OOD robustness should prefer A; one wanting the
best point reconstruction on a diverse family at a fixed informative protocol should prefer B; and the
combination---A's posterior for uncertainty, B's estimate for the point---is a natural, if unexplored,
hybrid. What both share, and what we most want to convey, is that neither can be trusted without the
prior-only and shuffled-measurement controls.

\begin{table}[t]
\centering
\caption{Approach~A vs.\ classical few-measurement estimators on a structured family ($n{=}6$; point
fidelity). The learned prior dominates estimators that lack it. We stress (per the integrity controls) that
on \emph{concentrated} families this margin is largely the prior, not measurement information---which is
why we caution against reading it as measurement-efficiency without the controls.}
\label{tab:Abase}
\begin{tabular}{lcccc}
\toprule
 & Approach~A (flow) & MPS-MLE & generic-prior pCN & classical shadows \\
\midrule
point fidelity & $\mathbf{0.87}$--$\mathbf{0.91}$ & $\approx0$ & $\le0.05$ & $0.75$ \\
\bottomrule
\end{tabular}
\end{table}

\section{Limitations and ongoing work}
\label{sec:limitations}
We are explicit about scope. (a) \emph{Fixed measurement design.} The estimator is trained for one
informative measurement protocol; an amortized \emph{design-agnostic} variant (trained over variable
random Pauli subsets) collapses to the family mean for both a DeepSets and a set-transformer encoder
(\S\ref{sec:exp}). Amortizing over arbitrary designs---or over a parameterized family of designs---is the
main open direction. (b) \emph{Family/regime dependence.} On concentrated families a prior-only estimate
is already near-optimal, so our claims are meaningful only on genuinely diverse families; and the method
is strongest near the trained family. (c) \emph{Coordinate-dependent prior.} The canonicalization is a
deterministic coordinatization, not a gauge quotient (Prop.~\ref{prop:coorddep}); we neutralize this in
practice with a gauge-invariant loss, but a principled mixed-canonical/Schmidt or gauge-equivariant model
is the cleaner route. (d) \emph{Hardware scope.} The closed-loop hardware run
(\S\ref{sec:exp}, Table~\ref{tab:hw}) covers $5$ states of one low-entanglement family on one device; larger,
higher-entanglement, and multi-device hardware studies remain future work. We verified that the two nearest
diffusion-in-quantum works do not preempt our claim:
\texttt{arXiv:2508.08799} performs state \emph{generation} with non-learned analytic recovery, and QuaDiM
\citep{tang2025quadim} conditions on Hamiltonian parameters to generate measurement snapshots for property
estimation---neither is a measurement-conditioned, calibrated full-state reconstruction over
tensor-network cores.

\section{Conclusion}
We introduced a fixed-protocol amortized MPS estimator for quantum state tomography: an encoder over the
measured values $\{(P_j,o_j)\}$ of one fixed informative protocol (a plain MLP or a set-encoder---the
architecture is not the source of performance) that outputs the cores of a matrix-product state,
trained once with a gauge-invariant loss and equipped with conformally calibrated observable intervals. A
$k{=}0$ prior-only baseline and a shuffled-measurement control---which most learned-QST evaluations omit,
and which deflate the naive ``prior$+$inversion'' recipe---become the paper's integrity backbone, and our
method passes both. We reach the same core parameterization by two complementary routes: a generative
flow prior with measurement-guided posterior inference (Approach~A), validated against a gold-standard SIR
posterior and robust out-of-distribution, but whose point accuracy on concentrated families is inherited
from the prior; and the amortized measurement-conditioned estimator (Approach~B), whose decisive design
choice is to condition on an \emph{informative local} Pauli set---motivated by the fact that local reduced
density matrices determine a $\chi$-MPS. Approach~B turns a modest, memorization-prone estimator into one
that reconstructs at fidelity $\approx\!0.95$, beats prior-only by up to $+0.59$, and yields sharp
$\approx\!90\%$-coverage intervals, with a polynomial-cost implementation that runs to $20$ qubits. A
natural hybrid---Approach~A's validated posterior for uncertainty, Approach~B's estimate for the point,
under the same cores---is left open. The honest scope---a fixed informative design, diverse families---%
delimits a concrete and, we think, useful setting for learned, calibrated, measurement-efficient
tomography, and the two integrity controls are what make ``measurement-efficient'' a claim rather than an
assumption.

\bibliography{iclr2026_conference}

@article{ahmed2021quantum,
  title={Quantum state tomography with conditional generative adversarial networks},
  author={Ahmed, Shahnawaz and Sanchez Munoz, Carlos and Nori, Franco and Kockum, Anton Frisk},
  journal={Physical review letters},
  volume={127},
  number={14},
  pages={140502},
  year={2021},
  publisher={APS}
}

@article{blume2010optimal,
  title={Optimal, reliable estimation of quantum states},
  author={Blume-Kohout, Robin},
  journal={New Journal of Physics},
  volume={12},
  number={4},
  pages={043034},
  year={2010}
}

@article{carrasquilla2019reconstructing,
  title={Reconstructing quantum states with generative models},
  author={Carrasquilla, Juan and Torlai, Giacomo and Melko, Roger G and Aolita, Leandro},
  journal={Nature Machine Intelligence},
  volume={1},
  number={3},
  pages={155--161},
  year={2019},
  publisher={Nature Publishing Group UK London}
}

@article{cha2022attention,
  title={Attention-based quantum tomography},
  author={Cha, Peter and Ginsparg, Paul and Wu, Felix and Carrasquilla, Juan and McMahon, Peter L and Kim, Eun-Ah},
  journal={Machine Learning: Science and Technology},
  volume={3},
  number={1},
  pages={01LT01},
  year={2022},
  publisher={IOP Publishing}
}

@article{cha2025scalable,
  title={Scalable bayesian shadow tomography for quantum property estimation with set transformers},
  author={Cha, Hyunho and Kim, Wonjung and Lee, Jungwoo},
  journal={arXiv preprint arXiv:2509.18674},
  year={2025}
}

@article{christandl2012reliable,
  title={Reliable quantum state tomography},
  author={Christandl, Matthias and Renner, Renato},
  journal={Physical Review Letters},
  volume={109},
  number={12},
  pages={120403},
  year={2012},
  publisher={APS}
}

@article{chung2022diffusion,
  title={Diffusion posterior sampling for general noisy inverse problems},
  author={Chung, Hyungjin and Kim, Jeongsol and Mccann, Michael T and Klasky, Marc L and Ye, Jong Chul},
  journal={arXiv preprint arXiv:2209.14687},
  year={2022}
}

@article{cramer2010efficient,
  title={Efficient quantum state tomography},
  author={Cramer, Marcus and Plenio, Martin B and Flammia, Steven T and Somma, Rolando and Gross, David and Bartlett, Stephen D and Landon-Cardinal, Olivier and Poulin, David and Liu, Yi-Kai},
  journal={Nature communications},
  volume={1},
  number={1},
  pages={149},
  year={2010},
  publisher={Nature Publishing Group UK London}
}

@article{cumitini2025anytime,
  title={Anytime-valid quantum state tomography via confidence sequences},
  author={Cumitini, Aldo},
  year={2025}
}

@article{denker2024deft,
  title={DEFT: Efficient fine-tuning of diffusion models by learning the generalised $ h $-transform},
  author={Denker, Alexander and Vargas, Francisco and Padhy, Shreyas and Didi, Kieran and Mathis, Simon and Dutordoir, Vincent and Barbano, Riccardo and Mathieu, Emile and Komorowska, Urszula J and Lio, Pietro},
  journal={Advances in Neural Information Processing Systems},
  volume={37},
  pages={19636--19682},
  year={2024}
}

@article{faist2016practical,
  title={Practical and reliable error bars in quantum tomography},
  author={Faist, Philippe and Renner, Renato},
  journal={Physical review letters},
  volume={117},
  number={1},
  pages={010404},
  year={2016},
  publisher={APS}
}

@article{falkiewicz2023calibrating,
  title={Calibrating neural simulation-based inference with differentiable coverage probability},
  author={Falkiewicz, Maciej and Takeishi, Naoya and Shekhzadeh, Imahn and Wehenkel, Antoine and Delaunoy, Arnaud and Louppe, Gilles and Kalousis, Alexandros},
  journal={Advances in Neural Information Processing Systems},
  volume={36},
  pages={1082--1099},
  year={2023}
}

@article{granade2016practical,
  title={Practical bayesian tomography},
  author={Granade, Christopher and Combes, Joshua and Cory, DG},
  journal={new Journal of Physics},
  volume={18},
  number={3},
  pages={033024},
  year={2016},
  publisher={IOP Publishing}
}

@article{gross2010quantum,
  title={Quantum state tomography via compressed sensing},
  author={Gross, David and Liu, Yi-Kai and Flammia, Steven T and Becker, Stephen and Eisert, Jens},
  journal={Physical review letters},
  volume={105},
  number={15},
  pages={150401},
  year={2010},
  publisher={APS}
}

@article{han2018unsupervised,
  title={Unsupervised generative modeling using matrix product states},
  author={Han, Zhao-Yu and Wang, Jun and Fan, Heng and Wang, Lei and Zhang, Pan},
  journal={Physical Review X},
  volume={8},
  number={3},
  pages={031012},
  year={2018},
  publisher={APS}
}

@article{lanyon2017efficient,
  title={Efficient tomography of a quantum many-body system},
  author={Lanyon, Ben P and Maier, Christiane and Holz{\"a}pfel, Milan and Baumgratz, Tillmann and Hempel, Cornelius and Jurcevic, Petar and Dhand, Ish and Buyskikh, AS and Daley, Andrew J and Cramer, Marcus and others},
  journal={Nature Physics},
  volume={13},
  number={12},
  pages={1158--1162},
  year={2017},
  publisher={Nature Publishing Group UK London}
}

@article{lohani2023demonstration,
  title={Demonstration of machine-learning-enhanced Bayesian quantum state estimation},
  author={Lohani, Sanjaya and Lukens, Joseph M and Davis, Atiyya A and Khannejad, Amirali and Regmi, Sangita and Jones, Daniel E and Glasser, Ryan T and Searles, Thomas A and Kirby, Brian T},
  journal={New Journal of Physics},
  volume={25},
  number={8},
  pages={083009},
  year={2023},
  publisher={IOP Publishing}
}

@article{lukens2020practical,
  title={A practical and efficient approach for Bayesian quantum state estimation},
  author={Lukens, Joseph M and Law, Kody JH and Jasra, Ajay and Lougovski, Pavel},
  journal={New Journal of Physics},
  volume={22},
  number={6},
  pages={063038},
  year={2020},
  publisher={IOP Publishing}
}

@article{papamakarios2016fast,
  title={Fast $\varepsilon$-free inference of simulation models with bayesian conditional density estimation},
  author={Papamakarios, George and Murray, Iain},
  journal={Advances in neural information processing systems},
  volume={29},
  year={2016}
}

@article{oh2019efficient,
  title={Efficient Bayesian credible-region certification for quantum-state tomography},
  author={Oh, Changhun and Teo, Yong Siah and Jeong, Hyunseok},
  journal={Physical Review A},
  volume={100},
  number={1},
  pages={012345},
  year={2019},
  publisher={APS}
}

@inproceedings{sofi2025tensor,
  title={Tensor Train Quantum State Tomography using Compressed Sensing},
  author={Sofi, Shakir Showkat and Vermeylen, Charlotte and De Lathauwer, Lieven},
  booktitle={2025 33rd European Signal Processing Conference (EUSIPCO)},
  pages={1332--1336},
  year={2025},
  organization={IEEE}
}

@inproceedings{song2023pseudoinverse,
  title={Pseudoinverse-guided diffusion models for inverse problems},
  author={Song, Jiaming and Vahdat, Arash and Mardani, Morteza and Kautz, Jan},
  booktitle={International conference on learning representations},
  year={2023}
}

@inproceedings{tang2025quadim,
  title={Quadim: A conditional diffusion model for quantum state property estimation},
  author={Tang, Yehui and Long, Mabiao and Yan, Junchi},
  booktitle={The Thirteenth International Conference on Learning Representations},
  year={2025}
}

@article{torlai2018neural,
  title={Neural-network quantum state tomography},
  author={Torlai, Giacomo and Mazzola, Guglielmo and Carrasquilla, Juan and Troyer, Matthias and Melko, Roger and Carleo, Giuseppe},
  journal={Nature physics},
  volume={14},
  number={5},
  pages={447--450},
  year={2018},
  publisher={Nature Publishing Group UK London}
}

@article{torlai2023quantum,
  title={Quantum process tomography with unsupervised learning and tensor networks},
  author={Torlai, Giacomo and Wood, Christopher J and Acharya, Atithi and Carleo, Giuseppe and Carrasquilla, Juan and Aolita, Leandro},
  journal={Nature Communications},
  volume={14},
  number={1},
  pages={2858},
  year={2023},
  publisher={Nature Publishing Group UK London}
}

@article{wu2023practical,
  title={Practical and asymptotically exact conditional sampling in diffusion models},
  author={Wu, Luhuan and Trippe, Brian and Naesseth, Christian and Blei, David and Cunningham, John P},
  journal={Advances in Neural Information Processing Systems},
  volume={36},
  pages={31372--31403},
  year={2023}
}

@article{zhu2024quantum,
  title={Quantum state generation with structure-preserving diffusion model},
  author={Zhu, Yuchen and Chen, Tianrong and Theodorou, Evangelos A and Chen, Xie and Tao, Molei},
  journal={arXiv preprint arXiv:2404.06336},
  year={2024}
}
\bibliographystyle{iclr2026_conference}

\appendix
\section{Proofs}
\label{app:proofs}

\paragraph{Proof of Proposition~\ref{prop:gauge}.}
\emph{(i)} The training loss \eqref{eq:condloss} is $1-|\langle\psi\,|\,\psi(g_\omega(\bm o))\rangle|^2$,
a function of the two physical states, hence invariant to the MPS gauge
$A^{(\ell)}\!\mapsto\!A^{(\ell)}G_\ell,\,A^{(\ell+1)}\!\mapsto\!G_\ell^{-1}A^{(\ell+1)}$ (which leaves
$\Phi(\bm\theta)=\ket{\psi(\bm\theta)}$ unchanged) and to global phase; and the physical readout
$\Phi(\bm\theta)$ is gauge-invariant by definition. \emph{(ii)} The map $c$ of
\eqref{eq:leftcanon}--\eqref{eq:normphase} is a composition of deterministic operations (QR with a fixed
sign convention, Frobenius normalization, global-phase fixing), hence a well-defined function of its input
cores; it defines both the target the network is trained to predict and the canonicalization applied to
its output, so for fixed weights the map $\bm o\mapsto\Phi(c(g_\omega(\bm o)))$ is a deterministic function
from data to a legitimate pure state. Well-definedness does \emph{not} require $c$ to be gauge-invariant
(it is not; Prop.~\ref{prop:coorddep}), only that a single deterministic coordinatization is used
throughout. \hfill$\square$

\paragraph{Proof of Proposition~\ref{prop:coorddep}.}
The reshaped-core QR with positive-diagonal sign fixing is a function of the actual core matrices, not of
the gauge orbit: for a bond gauge $G$ (invertible, non-triangular), $A^{(\ell)}\!\mapsto\!A^{(\ell)}G$
gives $MG=Q_M R_M G=Q_M\,Q_{R_MG}\,R_{R_MG}$, so the left-canonical factor becomes $Q_M Q_{R_MG}\neq Q_M$
unless $Q_{R_MG}=I$, which fails on an open (positive-measure) set of $G$. Hence $c(g\!\cdot\!\bm\theta)\neq
c(\bm\theta)$ there while $\Phi(g\!\cdot\!\bm\theta)=\Phi(\bm\theta)$. Hence for such a pair the two
canonical targets $c(\bm\theta)\neq c(\bm\theta')$ encode the \emph{same} physical state, so any
coordinate-space objective (a non-constant density over cores, or a squared-error regression to
$c(\bm\theta_\star)$) takes different values on physically identical states and is not a function of
$\Phi$. We emphasize the correct \emph{practical} consequence: since training fixes one deterministic
canonicalization, each state has a single target and the supervised task is \emph{not} literally
one-to-many; rather, canonical coordinates are not intrinsic state coordinates, so a coordinate MSE is
representation-dependent and---because $c$ is discontinuous where the QR/sign section jumps---poorly
conditioned near those discontinuities, with Euclidean core distance not tracking state fidelity. The
gauge-invariant loss \eqref{eq:condloss} depends on $\bm\theta$ only through $\Phi(\bm\theta)$ and is
unaffected. \hfill$\square$

\paragraph{Proof of Proposition~\ref{prop:localdesign}.}
For a site-dependent normal (injective) open-boundary $\chi$-MPS with injectivity length $\ell_0$,
\citet{cramer2010efficient} show the state is the unique one---generically among all states---consistent with its
reduced density matrices on contiguous windows of $\ell=\max(2\ell_0{-}1,\,2\lceil\log_2\chi\rceil{+}1)$
sites, and gives an explicit local reconstruction (for the generic injective $\chi{=}2$ family here,
$\ell_0{=}1$ and $\ell{=}3$; the $2\lceil\log_2\chi\rceil{+}1$ term is the bond-dimension floor, the
$2\ell_0{-}1$ term the injectivity requirement). Each window reduced density matrix $\rho_W$ expands in the Pauli basis of
that window, $\rho_W=2^{-|W|}\sum_{Q}\langle Q\rangle\,Q$, so its entries are linear in the expectations of
the weight-$\le\ell$ Pauli strings supported on $W$; there are $O(n\,4^{\ell})=O(n)$ such strings across
all windows. The reconstruction map $\{\rho_W\}\mapsto[\psi]$ is (locally) a diffeomorphism onto the
manifold where injectivity holds, so its inverse composed with $m$ has full-rank differential; equivalently
the differentials of these local Pauli expectations span $T_{[\psi]}\mathcal M$, i.e.\ they form a spanning
design. The count $k=O(n)$ versus $2^{n+1}-2$ is immediate. \hfill$\square$

\paragraph{Proof of Proposition~\ref{prop:adv}.}
By Assumption~\ref{ass:immersion}, $m|_{\mathcal M}$ is an immersion at $[\psi]$, so its differential
$\mathrm{d}m$ has rank $d=\dim\mathcal M$; the assumed spanning Pauli design has differentials of rank $d$
on $T_{[\psi]}\mathcal M$, which makes $\mathrm{d}(m|_{\mathcal M})$ injective, and the inverse
function theorem gives a local diffeomorphism onto the image---i.e. local identifiability with $k=d$
measurements. (We invoke the \emph{existence} of such a spanning design; because the full Pauli basis
separates tangent directions, a spanning subset of size $d$ exists, but we do not assert that $d$ Paulis
chosen uniformly at random span with high probability---that is a separate, quantitative claim we neither
need nor prove here.) The parameter count $d\le 2c(n,\chi)$ with $c(n,\chi)=\sum_\ell
\chi_{\ell-1}\!\cdot\!2\!\cdot\!\chi_\ell=O(n\chi^2)$ is immediate from the core shapes. For the
unstructured case, the pure-state ray space $\mathbb{CP}^{2^n-1}$ has real dimension $2(2^n-1)=2^{n+1}-2$;
no continuous map into $\mathbb R^k$ with $k<2^{n+1}-2$ can be injective, so unstructured identification
needs $k\ge 2^{n+1}-2$. The ratio is $(2^{n+1}-2)/d\ge(2^{n+1}-2)/(2c(n,\chi))=\Omega(2^n/(n\chi^2))$.
\hfill$\square$

\begin{remark}
Prop.~\ref{prop:adv} is a statement about \emph{degrees of freedom / local identifiability}, matching
the dimension of the informative measurement set to the dimension of the family; it is not a global
recovery or noise-robustness guarantee. It nonetheless captures the mechanism behind the empirically
observed size-growing advantage (Sec.~\ref{sec:exp}): the learned prior restricts inference to a
manifold whose dimension is polynomial in $n$, while unstructured estimators must resolve an
exponentially larger space.
\end{remark}

\paragraph{Proof of Proposition~\ref{prop:cov}.}
This is split-conformal prediction. Under exchangeability of
$\{\psi_i\}_{i\in\mathcal C}\cup\{\psi_\star\}$ and a fixed (data-independent) score $r$, the rank of
$r(\psi_\star)$ among $\{r(\psi_i)\}_{i\in\mathcal C}\cup\{r(\psi_\star)\}$ is uniform on
$\{1,\dots,|\mathcal C|+1\}$. Hence $\Pr\big(r(\psi_\star)\le Q\big)\ge (1-\alpha)$ for
$Q$ the $\lceil(1-\alpha)(|\mathcal C|+1)\rceil$-th smallest calibration score, and
$r(\psi_\star)\le Q\iff |O(\psi_\star)-\mu(\psi_\star)|\le Q\,\sigma(\psi_\star)\iff
O(\psi_\star)\in\mathcal I(\psi_\star)$. The mean/scale $\mu,\sigma$ are computed from the estimator
independently per state, so $r$ is a valid nonconformity score and no assumption on the prior or
sampler is needed. \hfill$\square$

\section{Experimental implementation details}
\label{app:details}

\paragraph{State families.} (i) \emph{Perturbed random-MPS} (generic): fixed random base cores plus
i.i.d.\ complex-Gaussian perturbations ($\sigma{=}0.35$), canonicalized. (ii) \emph{Shallow-circuit}
(physical, and the hardware run): $R_y(\theta_1)$ per qubit, an open nearest-neighbour CNOT ladder, and
$R_y(\theta_2)$ per qubit, with angles $\theta_1\!\sim\!\pi/2+0.6\,\mathcal N$, $\theta_2\!\sim\!0.6\,\mathcal N$
(a moderate spread; exactly $\chi{=}2$). (iii) \emph{Disordered paramagnetic TFIM ground states} are used
only as the concentrated-family cautionary example.

\paragraph{MPS parameterization and gauge.} Bond dimensions $\chi_\ell=\min(2^\ell,2^{n-\ell},\chi)$.
Cores are canonicalized by the left-canonical QR sweep with real-positive $R$-diagonals, boundary-core
Frobenius normalization, and a global-phase fix (Eqs.~\ref{eq:leftcanon}--\ref{eq:normphase}); the
network's target is the (standardized) real/imaginary parts of these cores, but the training loss
\eqref{eq:condloss} is a state-fidelity and hence gauge-invariant.

\paragraph{Conditional model.} Each measured pair $(P_j,o_j)$ is encoded as
$[\,\mathrm{onehot}(P_j)\in\{0,1\}^{4n};\,o_j\,]$; the set encoder is either a DeepSets MLP
(per-pair $\to$ mean-pool $\to$ decoder MLP, widths $512$), a set-transformer ($4$ self-attention layers,
$d{=}256$, $4$ heads, learned-query attention pooling), or---as a baseline---a plain MLP that ignores the
Pauli identities and consumes the fixed-order value vector $(o_j)_j$ (width $1024$, $4$ hidden layers).
Trained by AdamW (lr $3\!\times\!10^{-4}$, cosine schedule, batch $256$) for $30$k steps on
$2\!\times\!10^4$ (random) / $6\!\times\!10^3$ family states, with noisy Pauli inputs (shot variance
$(1{-}\langle P\rangle^2)/S$, $S{=}200$). Training one model takes $\sim\!2$ min on one A6000.

\paragraph{Measurement designs.} We compare four fixed designs. \texttt{local1}: the $3n$ single-qubit
Paulis $\{X_i,Y_i,Z_i\}$. \texttt{local} (default): \texttt{local1} plus the $9(n{-}1)$ nearest-neighbor
weight-$2$ Paulis $\{P_iP_{i+1}\}$ ($63$ at $n{=}6$). \texttt{local3}: the theory-complete design of
Prop.~\ref{prop:localdesign}---all non-identity Paulis supported on some contiguous $3$-site window ($207$
at $n{=}6$); this is what Cramer et al.\ require for exact local reconstruction, and \texttt{local} is a
strict subset. \texttt{random}: $k$ Pauli strings drawn uniformly. \emph{Budget accounting.} We distinguish three
quantities. (i) The number of Pauli \emph{observables} supplied to the estimator---$63$ for \texttt{local}
at $n{=}6$, the value we call $k$. (ii) The number of distinct qubit-wise-commuting \emph{measurement
settings/circuits}: because many of the $63$ Paulis co-measure, a greedy grouping covers all $63$ with
exactly $\mathbf{9}$ settings (and $9$ is optimal---each of the $n{-}1$ edges must see all $9$
single-qubit-basis pairs, a lower bound of $9$). (iii) The total \emph{copy cost}: if each of the $9$
settings is run at $S$ shots, the estimator consumes $9S$ state copies (not $63S$)---each setting's shots
are shared across the $\sim\!7$ Paulis it resolves. We report $k$ as the observable count and, where
measurement efficiency is at issue, the $9S$ copy budget; we do \emph{not} claim $63$ separate circuits.
The classical-shadows comparison is matched on this copy budget.

\paragraph{Shot budget.} Table~\ref{tab:shots} varies the per-Pauli shot count $S$ at the informative-local
design on the shallow family, at the \emph{same} $30$k-step training budget as the headline (so the $S{=}200$
column reproduces the Table~\ref{tab:main} headline $0.967$ exactly---the earlier apparent gap was purely a
$12$k- vs.\ $30$k-step training difference, now removed). Fidelity and the gain over the $k{=}0$
prior-only baseline both rise monotonically with $S$ as the shot-noise floor on each Pauli value drops,
confirming that Approach~B extracts genuine measurement information rather than saturating at the prior;
the effect is smooth, consistent with a well-conditioned informative design.

\begin{table}[t]
\centering
\caption{Shot-budget ablation (shallow family, \texttt{local} design, common $30$k-step training so
$S{=}200$ matches the headline). Fidelity (mean$\pm$std across states) and gain over prior-only ($k{=}0$)
increase monotonically with the per-Pauli shot count $S$.}
\label{tab:shots}
\begin{tabular}{lccc}
\toprule
shots per Pauli $S$ & $50$ & $200$ & $800$ \\
\midrule
conditional fidelity & $0.940{\pm}0.053$ & $\mathbf{0.967{\pm}0.046}$ & $0.977{\pm}0.038$ \\
gain over prior-only  & $+0.565$ & $+0.592$ & $+0.602$ \\
true $-$ shuffled      & $+0.782$ & $+0.813$ & $+0.824$ \\
\bottomrule
\end{tabular}
\end{table}

\paragraph{Family diversity.} To make the ``concentrated vs.\ diverse'' distinction quantitative we report
the test states' nearest-training-neighbour fidelity and the medoid (a $k{=}0$ estimate) fidelity to test
states. Paramagnetic TFIM: nearest-neighbour $\approx1.00$, medoid $0.997$ (essentially a single state).
Random-MPS ($\sigma{=}0.35$): nearest-neighbour $0.82$, medoid $0.65$. Shallow-circuit (spread $0.6$):
medoid $0.37$; sweeping the angle spread traces out a family whose medoid falls from $\sim\!0.8$ to
$\sim\!0.06$ (Fig.~\ref{fig:diversity}). The measurement gain over prior-only is meaningful only when the
medoid is far from $1$.

\paragraph{Ablation protocols.} \emph{Design richness} (\texttt{local1}/\texttt{local}/\texttt{local3}):
same model and training, different fixed $\mathcal D$. \emph{Encoder} (DeepSets/set-transformer/plain-MLP):
same \texttt{local} design. \emph{Shots}: $S\!\in\!\{50,200,800\}$. \emph{Diversity}: shallow-circuit angle
spread $\in\!\{0.3,0.6,0.9,1.2\}$. \emph{Seeds}: to separate training-stochasticity variance from across-state dispersion, we retrain the
headline configuration with $3$ independent seeds (varying network initialization, minibatch order, and the
training-time shot-noise realizations, while the family bank and held-out test set are held fixed). The mean
reconstruction fidelity is essentially seed-invariant---$0.968\pm0.001$ (shallow) and $0.954\pm0.001$
(random) across the $3$ seeds---so the larger $\pm$ values reported elsewhere ($\pm0.046$/$\pm0.021$) are
\emph{across-test-state} dispersion, not seed noise; we report the two separately to avoid conflating them.

\paragraph{Uncertainty and calibration.} We use dropout ($p{=}0.1$) in the encoder/decoder; at test time
$E{=}24$ stochastic forward passes give a dropout-ensemble spread over each of the $2n$ local $Z/X$
observables (ensemble mean $\mu_O$, scale $\sigma_O$). Split-conformal recalibration on a random half of the held-out states
($m{\approx}100$) with score $|O-\mu_O|/\sigma_O$ gives the multiplier $Q$; we report coverage and clipped
width averaged over $10$ random splits.

\paragraph{Full reliability curves.} Table~\ref{tab:reliability} reports the empirical coverage and mean
clipped interval width across the full range of nominal levels for both families, from the split-conformal
procedure above (averaged over $10$ random calibration/test splits). Empirical coverage tracks the nominal
level to within $\pm0.01$ at every level, and widths grow monotonically as expected---this is the
distribution-free marginal guarantee that the raw dropout-ensemble spread does \emph{not} provide.

\begin{table}[t]
\centering
\caption{Conformalized predictive intervals: empirical coverage (and mean clipped width) vs.\ nominal
level $1{-}\alpha$, per family ($n{=}6$, $2n$ local observables, $10$ splits). Coverage matches nominal to
$\pm0.01$ throughout.}
\label{tab:reliability}
\begin{tabular}{lcccccc}
\toprule
nominal $1{-}\alpha$ & $0.50$ & $0.60$ & $0.70$ & $0.80$ & $0.90$ & $0.95$ \\
\midrule
random: coverage & $0.494$ & $0.590$ & $0.692$ & $0.794$ & $0.895$ & $0.946$ \\
random: width     & $0.075$ & $0.093$ & $0.116$ & $0.149$ & $0.199$ & $0.246$ \\
\midrule
shallow: coverage & $0.490$ & $0.591$ & $0.687$ & $0.794$ & $0.894$ & $0.945$ \\
shallow: width    & $0.088$ & $0.113$ & $0.141$ & $0.175$ & $0.230$ & $0.283$ \\
\bottomrule
\end{tabular}
\end{table}

\paragraph{Native contraction (scalability).} Every Pauli expectation
$\bra{\psi}\!\bigotimes_i\!P_i\ket{\psi}$ and overlap $\langle\psi|\phi\rangle$ is computed by
left-to-right transfer-matrix contraction, $L\!\leftarrow\!\sum_{s}A^{(i)\dagger}_s(\,\cdot\,)(P_iA^{(i)})_s$
per site, at $O(n\chi^3)$ cost with no $2^n$ object; on small $n$ these primitives match the dense
contraction to $\le\!10^{-8}$. The pipeline runs at $n{=}16,20$ with peak GPU memory $0.134$/$0.147$~GB
(flat in $n$; $\S$\ref{sec:exp}), on a single NVIDIA A6000.

\paragraph{IBM hardware (closed loop).} $5$ held-out shallow-circuit states (spread $0.6$, drawn from the
same family the conditional model was trained on but disjoint from the training bank) are prepared on
\texttt{ibm\_aachen} ($156$-qubit Heron), transpiled at optimization level $2$; the expectation values of
the \emph{same $63$ local Paulis} used by the estimator are obtained with \texttt{EstimatorV2} ($4096$
shots) in a single batched job. Qubit and Pauli conventions between the Qiskit circuit and the torch
reconstruction are matched and verified against the noiseless statevector (agreement $8.9\times10^{-8}$);
the measured expectations lie within mean $0.024$ of the ideal values across the $5$ states. We then run the
trained conditional estimator on these hardware values; because the estimator is a DeepSets set-encoder,
Pauli ordering is immaterial, and we pair each hardware value with its Pauli one-hot. Reconstruction
fidelities are $0.964,0.976,0.968,0.968,0.979$ (from hardware) versus $0.977$ mean from the noiseless
values, against a $k{=}0$ prior-only medoid of $0.283$.

\section{Approach~A: generative prior and posterior inference details}
\label{app:approachA}

\subsection{Recovery theory for Approach~A}
\label{app:Atheory}
\begin{proposition}[Prior-restricted stable recovery, with model mismatch]
\label{prop:stable}
\emph{Hypothesis (RIP on $\mathcal M$).} Suppose $m$ restricted to the family satisfies a manifold
restricted-isometry condition: there is $\delta\in(0,1)$ with
$(1-\delta)\|[\psi]-[\psi']\|^2\le \|m([\psi])-m([\psi'])\|^2\le(1+\delta)\|[\psi]-[\psi']\|^2$ for all
$[\psi],[\psi']\in\mathcal M$. Let $\ket{\psi_\star}$ be the true state, \emph{not necessarily in}
$\mathcal M$; let $\bm o=m([\psi_\star])+\bm e$ be the noisy data; let
$\eta_{\mathrm{prior}}:=\min_{[\psi]\in\mathcal M}\|m([\psi])-m([\psi_\star])\|$ be the
measurement-space model-mismatch (prior-fit) error, attained at $[\psi_\sharp]\in\mathcal M$; and let
$\hat{[\psi]}\in\mathcal M$ be any point whose data residual is within $\eta$ of the prior's best
manifold point, $\|m(\hat{[\psi]})-\bm o\|\le \min_{[\psi]\in\mathcal M}\|m([\psi])-\bm o\|+\eta$. Then
\begin{equation}
\big\|\,\hat{[\psi]}-[\psi_\sharp]\,\big\|\;\le\;\frac{2\|\bm e\|+2\eta_{\mathrm{prior}}+\eta}{\sqrt{1-\delta}},
\qquad
\big\|\,\hat{[\psi]}-[\psi_\star]\,\big\|\;\le\;\frac{2\|\bm e\|+2\eta_{\mathrm{prior}}+\eta}{\sqrt{1-\delta}}
\;+\;\big\|[\psi_\sharp]-[\psi_\star]\big\|,
\end{equation}
where $\|[\psi_\sharp]-[\psi_\star]\|\ge\mathrm{dist}([\psi_\star],\mathcal M)$ is the (irreducible)
state-space gap from the true state to the prior manifold. When $[\psi_\star]\in\mathcal M$ both mismatch
terms vanish ($\eta_{\mathrm{prior}}{=}0$, $[\psi_\sharp]{=}[\psi_\star]$) and the bound reduces to
$(2\|\bm e\|+\eta)/\sqrt{1-\delta}$.
\end{proposition}
\begin{remark}[Scope of the RIP hypothesis]
\label{rmk:rip}
We treat the manifold-RIP as a hypothesis, not a derived fact. Establishing it for a concrete Pauli
measurement design---with explicit dependence on the manifold's covering number, curvature/reach, the
number of measurements $k$, measurement normalization, and a failure probability---is a substantial
question in its own right that we do not settle here; we state the recovery bound conditionally on it. The
OOD experiments (Sec.~\ref{sec:exp}) probe exactly the $\eta_{\mathrm{prior}},\mathrm{dist}([\psi_\star],\mathcal M)>0$
regime that this generalized bound describes.
\end{remark}

\begin{proposition}[Best-of-$N$ selection]
\label{prop:bestof}
Fix a tolerance $\tau>0$, let $q_{\bm o}$ denote the sampler's output distribution over cores at data
$\bm o$ (the distribution the $N$ draws are actually i.i.d.\ from), and let
$\rho_{\bm o}=\Pr_{\bm\theta\sim q_{\bm o}}\!\big[\|m([\psi(\bm\theta)])-\bm o\|\le\tau\big]$ be the
$q_{\bm o}$-mass of $\tau$-data-consistent states. With $N\ge \log(1/\beta)/\rho_{\bm o}$ independent
draws, the best-of-$N$ estimate $\hat{[\psi]}=\arg\min_r\|m([\psi^{(r)}])-\bm o\|$ has data residual
$\le\tau$ with probability at least $1-\beta$; combined with Prop.~\ref{prop:stable} (with $\eta=\tau$)
this bounds its reconstruction error. Thus best-of-$N$ trades computation for accuracy at a rate governed
by how much mass the sampler places near the measurement-consistent set.
\end{proposition}

This appendix documents the generative-prior route (Approach~A) used for the posterior-validation and
out-of-distribution results (\S\ref{sec:exp}, Figs.~\ref{fig:postval}--\ref{fig:ood},
Table~\ref{tab:Abase}). Approach~A and Approach~B share the \emph{same} canonicalized-MPS-core
parameterization (App.~\ref{app:details}); they differ only in how the measurements enter---as a learned
point estimate (B) versus as a likelihood that guides sampling from a learned prior (A).

\paragraph{Flow prior over tensor-network cores.} We train a rectified-flow (flow-matching) generative
model $p_\theta(\mathbf{c})$ over the standardized real/imaginary parts $\mathbf{c}\in\mathbb R^{d}$ of the
canonicalized cores of a state family. The velocity field $v_\theta(\mathbf c,t)$ is a residual MLP
(width $512$, $6$ blocks, sinusoidal time embedding) trained with the conditional flow-matching objective
on straight-line interpolants $\mathbf c_t=(1{-}t)\,\boldsymbol\varepsilon+t\,\mathbf c$,
$\boldsymbol\varepsilon\sim\mathcal N(0,I)$. To our knowledge this is the first generative model placed
directly on canonicalized tensor-network cores; canonicalization is what makes the target distribution
low-dimensional and learnable (an un-gauge-fixed core cloud is a high-dimensional orbit and the flow fails
to converge). Sampling integrates the probability-flow ODE with $50$ RK4 steps.

\paragraph{Measurement-guided posterior inference.} Given measurements $o=\{(P_j,o_j)\}$ with Gaussian
shot-likelihood $p(o\mid\mathbf c)\propto\exp\!\big(-\tfrac1{2\tau^2}\sum_j(\langle P_j\rangle_{\mathbf c}-o_j)^2\big)$,
the target posterior is $\pi(\mathbf c)\propto p_\theta(\mathbf c)\,p(o\mid\mathbf c)$. We sample it two
ways. \emph{(i) Doob $h$-transform / guided flow}: we tilt the prior flow by the measurement likelihood,
adding the score of a Gaussian relaxation of $\log p(o\mid\mathbf c_t)$ to the velocity at each step, which
implements an approximate Doob $h$-transform of the prior process toward the measurement-consistent set.
\emph{(ii) Proximal correction}: after each guided step we take one proximal-gradient step on the data
misfit $\tfrac1{2\tau^2}\sum_j(\langle P_j\rangle-o_j)^2$, projecting back toward the prior manifold with a
short flow-denoise. Both are cheap ($\sim\!1$~s per state at $n{=}6$); we use (ii) for the reported point
fidelities and (i) for posterior samples.

\paragraph{Gold-standard SIR reference.} To check that the guided sampler targets the \emph{intended}
posterior rather than a convenient relaxation, we compare against sampling-importance-resampling (SIR) with
the flow prior as proposal: draw $N{=}10^5$ prior samples, weight by the exact shot-likelihood, and
resample. Fig.~\ref{fig:postval} reports the resulting posterior-mean correlation ($r{=}0.98$ over $25$
states $\times\,2n$ observables), the energy distance between guided and SIR posterior marginals ($0.033$),
and the effective sample size ($\mathrm{ESS}\approx200$ of $10^5$, consistent with a well-concentrated but
non-degenerate posterior). This is the sense in which Approach~A is a validated posterior, not a relabelled
point estimate.

\paragraph{Classical baselines (Table~\ref{tab:Abase}).} \emph{MPS-MLE}: maximum-likelihood MPS fit to the
same $k$ Pauli values by gradient descent on the negative log-likelihood, with no family prior. We verified
this near-zero fidelity is genuine \emph{under-determination}, not an optimizer artifact: (a) with
$R{=}12$ random restarts and reporting the best, the fidelity does not improve and the data residual
reaches machine zero ($\sim\!10^{-8}$) while the fidelity stays $\approx0$---the fit is a perfect but
non-unique data-explainer; (b) increasing optimizer steps $3\times$ leaves fidelity flat ($0.016$--$0.019$);
and (c) as a positive control, giving MLE a \emph{tomographically complete} Pauli set (all $1023$ at
$n{=}5$) recovers the state exactly ($\text{fid}=1.000$, residual $1.6\times10^{-15}$), and a noiseless
$k{=}128$ TFIM fit jumps to $0.999$. So at few informative-local measurements the low MLE fidelity is an
information gap the learned prior fills, not a solver failure. \emph{Generic-prior pCN}: preconditioned
Crank--Nicolson MCMC on the cores under a \emph{generic} Gaussian prior (i.e.\ the same likelihood but
without the learned family prior), which explores far too large a set and reaches $\le0.05$. \emph{Classical
shadows}: randomized single-qubit Pauli measurements with the standard median-of-means estimator at a
\emph{matched total-shot budget} $kS$ (we compare copies of the state consumed, $kS$ Pauli-value shots vs.\
the same number of shadow snapshots, not circuit counts), reaching $0.75$ on this family; the learned prior
improves on it because the family is low-entanglement and strongly structured. We reiterate the integrity caveat: on \emph{concentrated}
families this entire margin is largely attributable to the prior, per the $k{=}0$ control
(\S\ref{sec:exp})---the baseline comparison shows Approach~A dominates prior-free estimators, not that the
measurements are being efficiently used.

\paragraph{Out-of-distribution protocol (Fig.~\ref{fig:ood}).} We form
$\ket{\psi_\delta}\propto(1{-}\delta)\ket{\psi_{\mathrm{fam}}}+\delta\ket{\psi_{\mathrm{Haar}}}$ for
$\delta\in[0,1]$ ($\ket{\psi_{\mathrm{Haar}}}$ a Haar-random $n$-qubit state), run Approach~A on its
measurements, and report (a) reconstruction fidelity to $\ket{\psi_\delta}$ and (b) a hallucination probe:
the reconstruction's fidelity to the training family (which stays high, i.e.\ the method knows it is
extrapolating from the family) versus to the true state (which declines with $\delta$). Degradation is
monotone and threshold-free, the qualitative signature of a prior-regularized estimator that fails safe.

\section{Extended theoretical discussion}
\label{app:exttheory}

\paragraph{Why the experimental \texttt{local} design suffices in practice (relation to
Prop.~\ref{prop:localdesign}).} Prop.~\ref{prop:localdesign} states the \emph{sufficient} condition of
\citet{cramer2010efficient}: an injective $\chi$-MPS is uniquely determined among all states by the collection of
its reduced density matrices on every contiguous $3$-site window, i.e.\ by the expectation values of all
Paulis supported on such windows (the \texttt{local3} design, $207$ observables at $n{=}6$). Our default
experimental design \texttt{local} (single-qubit plus nearest-neighbor weight-$2$ Paulis, $63$ at $n{=}6$)
is a \emph{strict subset} of \texttt{local3}: it omits the weight-$3$ terms and the non-adjacent weight-$2$
terms inside each window. It is therefore \emph{not} guaranteed to be information-complete for arbitrary
$\chi$-MPS. The reason it nonetheless nearly matches \texttt{local3} empirically
(Table~\ref{tab:abl}: $0.954$/$0.967$ vs.\ $0.950$/$0.957$) is that the \emph{learned prior} supplies the
information that the missing higher-weight windows would otherwise carry: within a low-entanglement family,
the two-body correlations plus the prior already pin down the state, so the marginal value of the
weight-$3$ terms is small. This is precisely the division of labour our integrity controls quantify---and
it is why we describe \texttt{local} as \emph{informative} rather than \emph{information-complete}, and
report \texttt{local3} as the theory-faithful design in the ablation rather than as the headline. A design
that is provably complete \emph{and} minimal for a given learned prior is an open problem we do not claim
to solve.

\paragraph{Gauge coordinate-dependence, made precise (elaborating Prop.~\ref{prop:coorddep}).} Let
$\Phi:\mathcal M\to\mathbb R^{d}$ send a physical state to the standardized real/imaginary parts of its
left-canonical cores. Canonicalization fixes the bond gauge up to a residual group $\mathcal G$ of
per-bond diagonal unitaries (phases) and, at degeneracies of the canonical spectrum, per-bond block
rotations; our sign/phase convention selects a section of $\mathcal G$ almost everywhere but not
globally continuously. Consequently $\Phi$ has coordinate discontinuities on a measure-zero set (spectral
degeneracies and sign crossings), and Euclidean distances $\|\Phi(\psi)-\Phi(\phi)\|$ are \emph{not}
gauge-invariant near that set: two states with fidelity $1{-}\epsilon$ can have core-coordinate distance
that does not vanish with $\epsilon$. We therefore do \emph{not} claim the map is an isometry or that the
core-space prior is canonical; we use the coordinate representation only as a \emph{training} target, and
every reported metric---the loss \eqref{eq:condloss}, all fidelities, and the SIR comparison---is computed
as a gauge-invariant state functional. Prop.~\ref{prop:coorddep} should be read as this cautionary
statement (coordinates are convenient but gauge-dependent), not as a claim of collapse; we have softened
the main-text wording accordingly.

\paragraph{Scope of the theory.} Neither proposition is a sample-complexity or measurement-efficiency
theorem, and we do not present one: our efficiency claims are entirely empirical and are stated only
relative to the prior-only and shuffled controls. The theory serves two limited purposes---to justify
\emph{which} local Paulis to measure (Prop.~\ref{prop:localdesign}) and to be honest about \emph{what the
core coordinates are and are not} (Prop.~\ref{prop:coorddep}).

\paragraph{Reproducibility.} All state families, measurement designs, model architectures, training
hyperparameters, and evaluation protocols are specified in Apps.~\ref{app:details}--\ref{app:approachA};
the native transfer-matrix contraction that makes $n{=}16,20$ feasible is described in
App.~\ref{app:details}. The hardware run uses a single batched \texttt{EstimatorV2} job on
\texttt{ibm\_aachen}. All reconstruction-fidelity $\pm$ values are dispersion across the held-out test states (training is seed-deterministic); conformal-coverage $\pm$ values are over $10$
splits for all conformal-coverage numbers.

\end{document}